\documentclass[12pt]{article} 
\usepackage{natbib}
\usepackage{array,epsfig,fancyheadings,rotating}
\usepackage{sectsty, secdot}

\usepackage{amsmath}
\usepackage{amssymb}
\usepackage{amsfonts}
\usepackage{multirow}
\usepackage{amsthm}
\usepackage{url}

\newtheorem{theorem}{Theorem}

\theoremstyle{definition}

\newtheorem{example}{Example}

\usepackage{comment}
\usepackage{mathrsfs}
\usepackage{color}
\usepackage{pgfplots}
\usepackage{filecontents}
\usepackage{subfigure}
\usepackage{paralist}
\usepackage{bm}
\usepackage{booktabs}
\usepackage{multirow}

\usepackage{algorithm}
\usepackage{algorithmic}

\graphicspath{{figs_new/}}  

\allowdisplaybreaks[4] 

\newcommand{\calI}{{\cal I}}

\newcommand{\ie}{i.e., }

\newcommand{\calL}{{\cal L}} 
\newcommand{\calH}{{\cal H}} 
\newcommand{\calD}{{\cal D}} 
\newcommand{\calx}{{\cal X}} 
\newcommand{\caly}{{\cal Y}} 
\newcommand{\cals}{{\cal S}} 
\newcommand{\calK}{{\cal K}} 
\newcommand{\calF}{{\cal F}} 
\newcommand{\calG}{{\cal G}} 
 
\newcommand{\calA}{{\cal A}}

\newcommand{\s}[1]{^{(#1)}}

\newcommand{\CUT}[1]{}

\usepackage{setspace} 

\begin{document}
\setlength{\abovedisplayskip}{7pt} 
\setlength{\belowdisplayskip}{7pt} 



\markboth{\hfill{\footnotesize\rm  LAN AND HE} \hfill}
{\hfill {\footnotesize\rm MAXIMIN DESIGNS FOR MIXED-TYPE VARIABLES} \hfill}

\renewcommand{\thefootnote}{}
$\ $\par


\fontsize{12}{14pt plus.8pt minus .6pt}\selectfont \vspace{0.8pc}
\centerline{\large\bf MAXIMIN DISTANCE DESIGNS FOR MIXED  }
\vspace{2pt} 
\centerline{\large\bf CONTINUOUS, ORDINAL, AND BINARY }
\vspace{2pt} 
\centerline{\large\bf VARIABLES}
\vspace{.4cm} 
\centerline{Hui Lan and Xu He} 
\vspace{.4cm} 
\centerline{\it Beijing University of Technology and Chinese Academy of Sciences }
 \vspace{.55cm} \fontsize{9}{11.5pt plus.8pt minus.6pt}\selectfont


\begin{quotation}
\noindent {\it Abstract:}
Computer experiments are pivotal for modeling complex real-world systems. Maximizing information extraction and ensuring accurate surrogate modeling necessitates space-filling designs, where design points extensively cover the input domain. While substantial research has been conducted on maximin distance designs for continuous variables, which aim to maximize the minimum distance between points, methods accommodating mixed-variable types remain underdeveloped. This paper introduces the first general methodology for constructing maximin distance designs integrating continuous, ordinal, and binary variables. This approach allows flexibility in the number of runs, the mix of variable types, and the granularity of levels for ordinal variables. We propose three advanced algorithms, each rigorously supported by theoretical frameworks, that are computationally efficient and scalable. Our numerical evaluations demonstrate that our methods significantly outperform existing techniques in achieving greater separation distances across design points.

\vspace{9pt}
\par
\end{quotation}\par

\def\thefigure{\arabic{figure}}
\def\thetable{\arabic{table}}

\renewcommand{\theequation}{\thesection.\arabic{equation}}

\fontsize{12}{14pt plus.8pt minus .6pt}\selectfont

\section{Introduction}
\label{sec:intro}

Computer simulations have emerged as powerful tools for understanding real-world systems~\citep{Santner:book}. 
To maximize information gained from the computer trials and facilitate accurate surrogate modeling, 
it is widely believed that design points should be spread throughout the entire input space, a concept known as ``space-filling''.
Consequently, an overwhelmingly used criterion to discriminate designs for computer experiments is the $L_2$ separation distance, 
defined as the minimal Euclidean distance between pairs of design points, 
\( \rho(\calD) = \min_{\bm{x}, \bm{y} \in \calD} \|\bm{x}-\bm{y}\|_2. \) 
A design is termed a maximin distance design if it 
possesses the highest possible separation distance among all designs with the same input space and size $n$~\citep{Johnson:1990}. In many studies, a design is also referred to as a maximin distance design 
if existing methods do not produce a better design. 
Maximin distance designs have been demonstrated to attain asymptotic D-optimality in Gaussian process emulation, effectively mitigate numerical errors, and exhibit robustness to simulation inaccuracies \citep{haaland2011accurate,haaland2018framework}. 

Considerable efforts have been devoted to constructing maximin distance designs when the input space $\calG$ is $[0,1]^p$, i.e., when all variables are continuous. 
Notably, thirty-three algorithmic searching methods have been developed to generate two-dimensional and three-dimensional maximin distance designs, as detailed on the website \url{http://www.packomania.com/}.
Algorithmic constructions for general $p$ have also been proposed by \citet{trosset1999approximate}, \citet{stinstra2003constrained}, and \citet{mu2017algorithmic}. 
Nevertheless, owing to the complexity of design optimization, it is computationally infeasible for algorithmic search methods to identify the optimal design when neither $n$ nor $p$ is small. 
To reduce the search space, one popular approach is to confine the search to Latin hypercube designs, which consist of the points $\{1/(2n), 3/(2n), \ldots, (2n-1)/(2n)\}$ in each dimension \citep{morris1995exploratory}.
Upon observing that maximin distance designs often exhibit a repetitive structure, \citet{he2019interleaved} proposed searching within interleaved lattice-based designs, identifying most of the contemporarily best designs in $p>3$.  
Finally, algebraic construction methods have been proposed by 
\citet{xiao2017construction}, \citet{wang2018optimal}, 
\citet{li2021method}, \citet{wang2022design}, and \citet{yuan2025construction},
among others. 
Although some designs constructed using algebraic methods have been proven to be optimal in terms of separation distance, the constraints of the mathematical tools used limit their applicability to general values of $n$ and $p$.

Although designs constructed for $\calG = [0,1]^p$ can be adapted to accommodate any input space of the form $\calG = \prod_{k=1}^p \calG_k = \prod_{k=1}^p [a_k, b_k]$, they are not suitable for computer experiments involving discretely-valued factors.
However, it is common for computer experiments to include ordinal or categorical variables.  
As an illustration, Magmasoft V4.4, a widely used software for simulating the alloy casting process, includes five types of grey cast iron materials, with their names and compositions listed in Table~\ref{tab:iron}. Clearly, the composition factor should be treated as an ordinal variable with five allowable levels.

  \begin{table}[t!] 
    \centering
    \caption{Composition (percent) of the five types of grey cast iron material.}     \label{tab:iron}\par
   \begin{tabular}{|cccccc|} \hline 
      Name & Carbon & Silicon & Manganese & Phosphorus & Sulfur\\ \hline
      GJL-150 & 3.55 & 2.35 & 0.65 & 0.20 & 0.1 \\ \hline
      GJL-200 & 3.35 & 2.05 & 0.70 & 0.20 & 0.1 \\ \hline
      GJL-250 & 3.25 & 2.05 & 0.70 & 0.15 & 0.1 \\ \hline
      GJL-300 & 3.20 & 1.60 & 0.80 & 0.10 & 0.1 \\ \hline
      GJL-350 & 3.00 & 1.40 & 0.90 & 0.10 & 0.1 \\ \hline
    \end{tabular}
    \end{table}
However, methods for constructing designs with ordinal inputs are limited.
One approach, sliced maximin distance design~\citep{ba2015optimal}, is suitable for computer experiments with one categorical variable and multiple continuous variables.
By ignoring the ordinal nature of the variable and treating it as categorical, this method can also be applied to problems involving  continuous variables and a single ordinal variable.
Another common technique constructs a design in $\calG = [0,1]^p$ and then rounds the values to the nearest allowable levels for ordinal variables~\citep{joseph2020designing}. However, both techniques may result in suboptimal designs. Additionally, \citet{sun2019synthesizing} proposed a method for constructing maximin distance designs when all variables are discrete. To the best of our knowledge, no method has yet been developed for generating maximin distance designs with general mixed continuous and ordinal variables.

Other than maximin distance designs, \citet{deng2015design} 
proposed marginally coupled designs applicable to problems with mixed continuous and categorical variables. 
However, the use of orthogonal arrays imposes constraints on the value of $n$ for these designs.
\citet{joseph2020designing} introduced a simulated annealing algorithm, akin to that of \citet{morris1995exploratory}, to construct designs with mixed continuous and discretely-valued variables that is optimal in a criterion reflecting projection properties of the design~\citep{Joseph2015maxpro}. 
While can be applied to problems with mixed continuous and ordianry variables, this algorithm is computationally inefficient for large designs, with all examples in the paper having $n < 50$. 
These designs usually have better projection properties than maximin distance designs and thus are more appealling when there are sparsity among variables and we do not know which variables are effecitve in predicting the outcome. 
On the other hand, maximin distance designs are useful when we have prior knowledge on variable importance or there is no sparsity.

\begin{figure}[!t]
	\begin{minipage}{1\linewidth}	
    \centering
		\subfigure[]{
			\label{fig:1}
			\includegraphics[width=0.38\linewidth,height=1.8in]{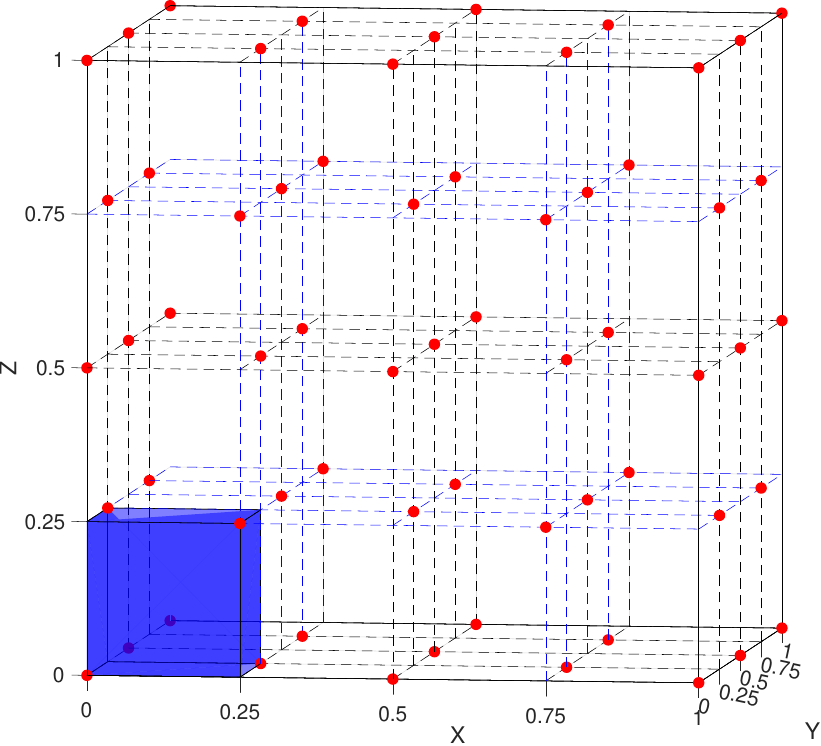}	
		}\noindent \hspace{1cm}
		\subfigure[]{
			\label{fig:2}
			\includegraphics[width=0.38\linewidth,height=1.8in]{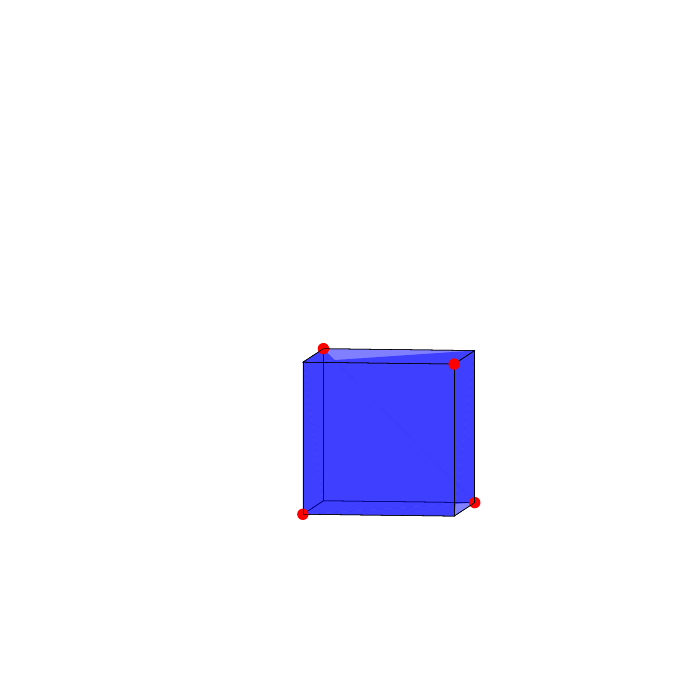}
		}
	\end{minipage}
  \vspace{2em}
	\begin{minipage}{1\linewidth }
    \centering
		\subfigure[]{
			\label{fig:3}
			\includegraphics[width=0.38\linewidth,height=1.8in]{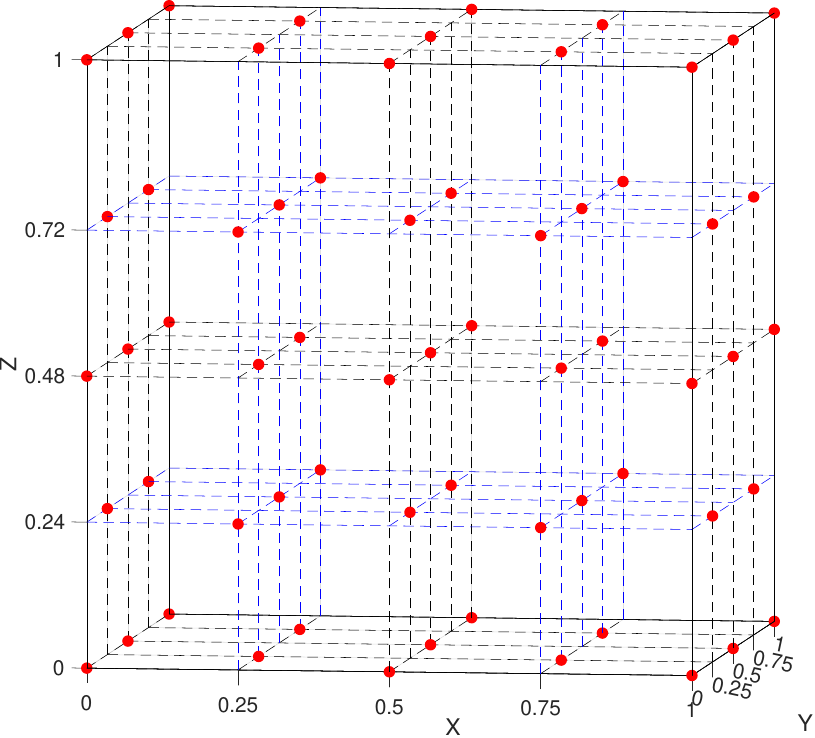}
		}\noindent \hspace{1cm}
		\subfigure[]{
			\label{fig:4}
			\includegraphics[width=0.38\linewidth,height=1.8in]{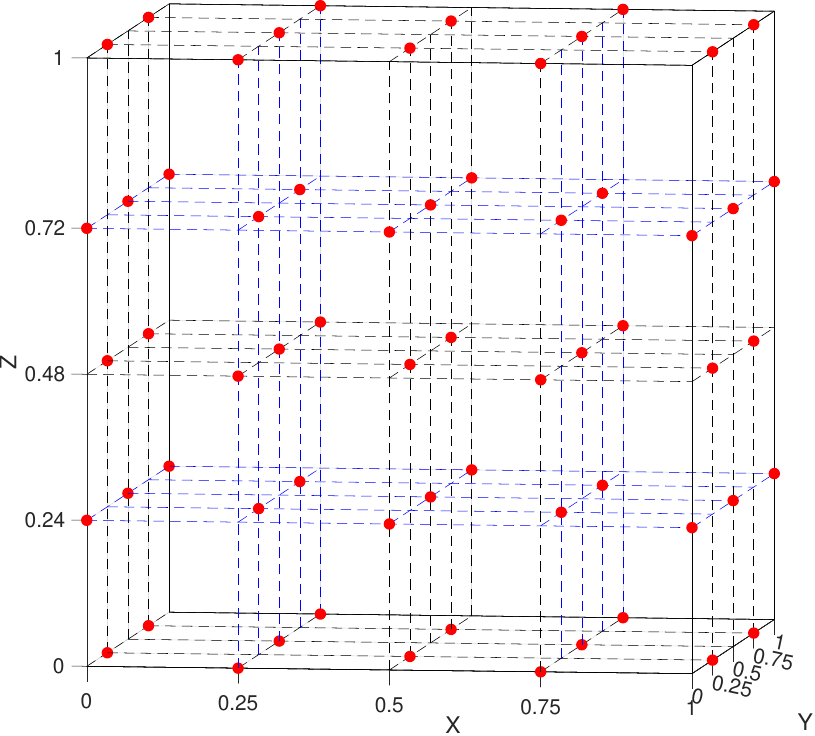}
		}
	\end{minipage}
  \vspace{-2em}
	\caption{
(a) An interleaved lattice-based maximin distance design for continuous variables  \citep{he2019interleaved} with $p=3$ and $n = 63$. 
(b) The local structure $\calH$ of the design in (a). 
(c) An interleaved lattice-based maximin distance design for mixed-type variables with $p=3$ and $n = 63$ for which $\bm{u}=(0,0,0)$.
(d) An interleaved lattice-based maximin distance design for mixed-type variables with $p=3$ and $n = 62$ for which $\bm{u}=(0,0,1)$. }
	\label{fig:example}
\end{figure}

In this paper, we propose the first general method to construct maximin distance designs with mixed continuous, ordinal, and binary variables,  offering flexibility in $n$, $p$, and the number of ordinal variables. 
We present three major construction algorithms, supported by theoretical results. 
Numerical results demonstrate that designs generated by our method achieve substantially greater separation distance than those produced by existing algorithmic approaches.
Furthermore, our methods are computationally efficient, enabling the construction of large designs.

Since binary variables are essentially ordinal with two levels, the main challenge is constructing designs with mixed continuous and ordinal variables.
We restrict the search to interleaved lattice-based designs, as proposed by \citet{he2019interleaved}, for four main reasons. First, unrestricted numerical search is computationally prohibitive for large-scale designs. Second, Latin hypercube designs are not suitable for problems involving ordinal variables, so we cannot impose the Latin hypercube constraint. Third, existing algebraic techniques for constructing designs in $[0,1]^p$ are not applicable to our problem.
Lastly but most importantly, 
the interleaved lattice constraint seems to be compatible to design spaces with discretely-valued variables. 
See Figure~\ref{fig:example} for an illustration. 
Panel (a) displays the maximin distance design obtained from \citet{he2019interleaved} for the continuous input space $\calG = [0,1]^3$ with $n = 63$.
It is referred to as having an interleaved lattice structure because the entire design is a translated repetition of the local structure shown in Panel (b).
Since the design is essentially a subset of the discrete space $\{0,0.25,0.5,0.75,1\}^3$, it is also a maximin distance design provided that the input space is $\{0,0.25,0.5,0.75,1\}^3$, $[0,1]^2 \times \{0,0.25,0.5,0.75,1\}$, or any other space containing $\{0,0.25,0.5,0.75,1\}^3$ as a subspace. 
Furthermore, after minor modifications, the design also qualifies as a maximin distance design for other discrete spaces. 
For instance, by replacing the 0.25, 0.5, and 0.75 in the first dimension with 0.24, 0.48, and 0.72, respectively, the design shown in Panel (c) becomes a maximin distance design for the space $\{0,0.24,0.48,0.72,1\} \times \{0,0.25,0.5,0.75,1\}^2$ or $\{0,0.24,0.48,0.72,1\} \times [0,1]^2$. 
To sum it, the techniques used to construct ordinary interleaved lattice-based designs can be extended to create 
designs with mixed types of variables.



\section{Preliminaries} \label{sec:pre}

In this section, we briefly review the construction method for interleaved lattice-based maximin distance designs for continuous variables~\citep{he2019interleaved}, referred to as ``ILMDC'' throughout this paper. 

The centeral problem for ILMDC is to find the design in $[0,1]^p$ that maximizes the separation distance, where the distance measure is $d(\bm{x},\bm{y}) = d(\bm{x}-\bm{y}) = \left\{ \sum_{k=1}^{p} (x_k - y_k)^2 \right\}^{1/2}$.
However, as discussed in \citet{he2019interleaved}, when prior knowledge about variable importance is available, 
the weighted $L_2$ distance \citep{ba2018sequential},  
$d_{\bm{w}}(\bm{x},\bm{y}) = d_{\bm{w}}(\bm{y}-\bm{x}) = \left\{ \sum_{k=1}^{p} \left\{w_k(x_k - y_k)\right\}^2 \right\}^{1/2}$, is preferred,  
in which the positive vector $\bm{w}= (w_1,\cdots,w_p)$, referred to as the weight vector, should be determined based on prior knowledge of variable importance.
In particular, variables with stronger impacts on the response should be assigned higher weights. 
For a set $\cals \subset \mathbb{R}^p$ and a vector $\bm{v} = (v_1,\ldots,v_p)$, let $\cals \otimes \bm{v}= \{ (x_1 v_1, \ldots, x_p v_p) : (x_1,\ldots,x_p) \in \cals \}$, which is a rescaling of $\cals$. 
According to \citet{he2019interleaved}, a design $\calD$ is the maximin distance design in $[0,1]^p$ under the distance measure $d_{\bm{w}}(\bm{x},\bm{y})$ if and only if the design $\calD \otimes \bm{w}$ is the maximin distance design in $\prod_{k=1}^p [0,w_k]$ using the standard distance measure $d(\bm{x}, \bm{y})$. 
Given this equivalence, this paper focuses on solving the latter problem.
Specifically, we employ the $L_2$ distance measure $d(\bm{x}, \bm{y})$ and assume that the input space for ILMDC is $\prod_{k=1}^p [0, w_k]$.

A set $\cals \subset \mathbb{R}^p$ is a lattice if it forms a group, i.e., $\bm{u} + \bm{v} \in \cals$ whenever $\bm{u}, \bm{v} \in \cals$. 
For any $p \times p$ matrix $\bm{G}$, the set $\calL(\bm{G})=\{\bm{z}^T \bm{G}: \bm{z} \in \mathbb{Z}^p \}$ forms a lattice, and  $\bm{G}$ is its generator matrix.
For instance, the $p$-dimensional integer lattice is $\mathbb{Z}^p = \calL(\bm{I}_{p\times p})$, where $\bm{I}_{p\times p}$ denotes the $p$-dimensional identity matrix. 
Another example is the $p$-dimensional even
integer lattice, $\mathbb{E}^p = \{ (x_1,\ldots,x_p) : x_k \text{ is an even integer for any } k \} = \calL(2\bm{I}_{p\times p})$.

A lattice $\calL$ is called a standard interleaved lattice (IL) if $\mathbb{E}^p \subset \calL \subset \mathbb{Z}^p$ and each dimension of $\calL$ takes any interger values. 
A comprehensive list of ILs is detailed in \citet{he2019interleaved}.
For a set $\cals \subset \mathbb{R}^p$ and a vector $\bm{u} \in \mathbb{R}^p$, let $\cals \oplus \bm{u}=\{ (x_1 + u_1, \ldots, x_p + u_p) : (x_1,\ldots,x_p) \in \cals \}$, which is a translation of $\cals$. 
A design $\calD$ in the space $\prod_{k=1}^p [0,w_k]$ is called an interleaved lattice-based design for continuous variables (ILDC) if it can be expressed as $\calD = \left\{ ( \calL \oplus \bm{u} ) \otimes \bm{v} \right\} \cap \prod_{k=1}^p [0,w_k]$,
\CUT{
\begin{equation}\label{eqn:ILDC}
 \calD = \left\{ ( \calL \oplus \bm{u} ) \otimes \bm{v} \right\} \cap \prod_{k=1}^p [0,w_k], 
\end{equation}
}
where $\calL$ is an IL, $\bm{v} \in (0,+\infty)^p$,
and $\bm{u} \in \mathbb{Z}^p$. 
In other words, the design $\calD$ consists of translated and rescaled lattice points located within the input space. 

Any IL $\calL$ can be expressed as $\calL = \cup_{\bm{z} \in \mathbb{Z}^p} \{ \calH \oplus (2\bm{z}) \}$, 
where $\calH = \calL \cap \{0,1\}^p$ is a finite set that determines $\calL$. 
Therefore, the $\calD$ can be viewed as the translated replication of an local point set $\calH \otimes \bm{v}$. 
For instance, Figure~\ref{fig:example}(a) illustrates a 3-dimensional ILMDC, which can be expressed as 
$\calD = \{ ( \calL \oplus \bm{u} ) \otimes \bm{v} \} \cap \prod_{k=1}^p [0,w_k]$, 
$\calL=\calL(\bm{G}_{3,2})$ in \eqref{eqn:G:3}, 
$\bm{u} = (0,0,0)$, $\bm{v} = (0.25,0.25,0.25)$,  and $\bm{w}= (1,1,1)$. 
It replicates the local point set shown in Figure~\ref{fig:example}(b). 
The design achieves a high separation distance $\rho(\calD)=0.3535$ because points within the same local set are well-spaced and no two points from different local sets are very close to each other.
To construct the ILMDC, we must select a local point set with excellent separation distance and determine how many times it should be replicated.

\citet{he2019interleaved} demonstrated that, to construct an ILMDC where the $k$th dimension has $s_k$ unique levels, it suffices to set $v_k = w_k / (s_k - 1)$ and $u_k \in \{0, 1\}$ since such choices lead to the optimal design. 
Furthremore, such designs can be expressed as:
\begin{equation}\label{eqn:ILDC:2}
 \calD = \left\{ ( \calL \oplus \bm{u} ) \cap \prod_{k=1}^p [0,s_k-1] \right\} \otimes \left(w_1/(s_1-1),\ldots,w_p/(s_p-1)\right).   
\end{equation}
Consequently, constructing the ILMDC reduces to searching for the optimal combination of $(\calL, \bm{s}, \bm{u})$. 
Remark that directly optimizing the design in $\prod_{k=1}^p [0,w_k]$ is a challenging problem unless both $n$ and $p$ are small because it is an $np$ dimensional optimization problem with the presence of numerous local optima.
However, by reducing the problem to optimizing $(\calL, \bm{s}, \bm{u})$ and employing several computational techniques we shall introduce, the optimization becomes significantly more manageable. 
This explains why most contemporarily maximin distance designs are ILMDCs~\citep{he2019interleaved}.

\section{Methodology for mixed-type variables} \label{sec:design}

\subsection{General expression}\label{sec:expression}

This section introduces the construction method for interleaved lattice-based maximin distance designs for mixed-type variables, hereafter referred to as ``ILMDM''.

We assume that $\calG_k$ represents the input space for the $k$th variable. For continuous variables, $\calG_k = [0, w_k]$, where $w_k$ reflects the variable's importance. For ordinal variables, $\calG_k$ is a finite subset of $[0, w_k]$ containing both 0 and $w_k$. For binary variables, $\calG_k = \{0, w_k\}$. The overall input space is $\calG = \prod_{k=1}^p \calG_k$.
The problem is formally defined as identifying the interleaved lattice-based design for mixed-type variables (ILDM), $\calD \subset \calG$, that maximizes the separation distance among all designs with a size of at least $n$, given $\calG$ and $n$.

We first present the foundational theory underlying ILDM, which streamlines the design construction process.
Firstly, recall that the levels of the $k$th dimension of an ILMDC given in 
\eqref{eqn:ILDC:2} are 
$\{0, w_k/(s_k-1), \ldots, w_k\}$
when $v_k \in (0,+\infty)$, $u_k \in \mathbb{Z}$, and $s_k = w_k/v_k +1$. 
Since the levels in $\calG_k$ may not be equally spaced, a new general formulation for ILDMs is needed.
Throughout this work, we consider ILDMs expressible as
\begin{equation}\label{eqn:ILDM}
 \calD = \left\{ (\caly_{1(i_1)},\ldots,\caly_{p(i_p)}) : (i_1,\ldots,i_p) \in \left\{ ( \calL \oplus \bm{u} ) \cap \prod_{k=1}^p [0,s_k-1] \right\} \oplus \bm{1}_p \right\}, 
\end{equation}
where $\bm{z}_p$ denotes the $p$-vector in which all entries are $z$, $\caly_{k(i)}$ denotes the $i$th smallest entry of $\caly_{k}$, and $\caly_k \subset \calG_k$ specifies the levels of the $k$th dimension of the design $\calD$. Additionally, $\caly = \prod_{k=1}^p \caly_k \subset \calG$ defines the overall level set of the design.
These designs generalize \eqref{eqn:ILDC:2} by allowing for flexible levels. If $\caly_k = \{ i w_k/(s_k-1) : i=0,1,\ldots,s-1 \} $ for any $k$, the design in \eqref{eqn:ILDM} reduces to the form in \eqref{eqn:ILDC:2}. 
From \eqref{eqn:ILDM}, the design $\calD$ is fully specified by $\calL$, $\caly$, $\bm{s}$, and $\bm{u}$. Henceforth, we denote the design as $\calD = \calD(\calL,\bm{s},\caly,\bm{u})$.

In particular, Figure~\ref{fig:example}(c) and (d) illustrate two ILDMs.  They share the same lattice structure $\calL=\calL(\bm{G}_{3,2})$, $\bm{s} = (5,5,5)$, $\caly=\{0,0.24,0.48,0.72,1\}\times \{0,0.25,0.5,0.75,1\}^2$, and $\rho(\calD)=0.3466$ is marginally lower than that of the design in Figure~\ref{fig:example}(a), which is $0.3535$.
Both designs 
qualify as ILMDM for various input spaces, including
$\calG = \{0,0.24,0.48,0.72,1\} \allowbreak \times \{0,0.25,0.5,0.75,1\}^2$ 
and $\calG = \{0,0.24,0.48,0.72,1\}\times [0,1]^2$. 
Differences in size between two designs result from variations in their respective $\bm{u}$ vectors.
Throughout this paper, $m(\cals)$ denotes the size of a design or set $\cals$.
The design in Figure~\ref{fig:example}(c) has $\bm{u}=(0,0,0)$ and $m=63$, whereas the design in Figure~\ref{fig:example}(d) has $\bm{u}=(0,0,1)$ and $m=62$. A brief reflection reveals that the size of an ILDM, $m\{\calD(\calL,\bm{s},\caly,\bm{u})\} = m\left\{ ( \calL \oplus \bm{u} ) \cap \prod_{k=1}^p [0, \allowbreak s_k - \allowbreak 1] \allowbreak \right\}$, depends solely on $\calL$, $\bm{s}$, and $\bm{u}$, and is independent of $\caly$.
For simplicity, we frequently abbreviate $m\{\calD(\calL,\bm{s},\caly,\bm{u})\}$ as $m(\calL,\bm{s},\bm{u})$. Furthermore, theoretical results regarding sample size for ILDC are equally applicable to ILDM. Theorems~\ref{thm:0_p:m}-\ref{thm:m:r} outline several important findings in this context.
Firstly, Theorem~\ref{thm:0_p:m} states that ILDMs with $\bm{u}=\bm{0}_p$ have the largest possible size among ILDMs of the same $\calL$ and $\bm{s}$. 

\begin{theorem} \label{thm:0_p:m}
Suppose 
$\calL$ is an IL, $\caly \subset \calG$, $\bm{s} \in \mathbb{N}^p$, $\bm{u} \in \mathbb{Z}^p$, and $\calD(\calL,\bm{s},\caly,\bm{u})$ is an ILDM in \eqref{eqn:ILDM}. 
Then $m\{\calD(\calL,\bm{s},\caly,\bm{u})\} \leq m\{\calD(\calL,\bm{s},\caly, \allowbreak \bm{0}_p)\}$.
\end{theorem}

Two essential quantities associated with an IL are its $q$ and $r$ values. 
Let $\bm{e}_k$ denote the unit $p$-vector whose $k$th entry is one and other entries are zero. 
Then, $q(\calL)=\log_2 \{m(\calH)\}$ represents the logarithm of the number of points in the lattice within $\{0,1\}^p$, and $r(\calL)=m(\{k :\bm{e}_k \in \calL\})$ specifies the number of standard basis vectors contained in the lattice.
The integers $q$ and $r$ must satisfy either $r = q = p$ or $0 \le r < q < p$~\citep{he2017interleaved}. 
The size of an ILDM is predominantly determined by $q$ and $\bm{s}$,  with only weak dependence on other attibutes of $\calL$, as established in Theorems~\ref{thm:m:s} and~\ref{thm:m:r}. 

\begin{theorem} \label{thm:m:s}
Suppose 
$\calL$ is an IL, $\caly \subset \calG$, $\bm{s} \in \mathbb{N}^p$, $\bm{u} \in \mathbb{Z}^p$, and $\calD(\calL,\bm{s},\caly,\bm{u})$ is an ILDM in \eqref{eqn:ILDM}. 
Then 
$2^{q-p} \prod_{\bm{e}_k \notin \calL} (2 \lfloor s_k/2 \rfloor ) \prod_{\bm{e}_k \in \calL} s_k \le m\{\calD(\calL,\bm{s},\caly, \allowbreak \bm{u})\} \le 2^{q-p} \prod_{\bm{e}_k \notin \calL} (2 \lceil s_k/2 \rceil ) \prod_{\bm{e}_k \in \calL} s_k.$
\end{theorem}

\begin{theorem} \label{thm:m:r}
For any $\bm{s} \in \mathbb{N}^p$ and $0\le z_1 \le z_2 < z_3<p$, 
 $\max_{q(\calL)=z_3,r(\calL)=z_2} \\m(\calL,\bm{s},\bm{u}) \ge \max_{q(\calL)=z_3,r(\calL)=z_1}m(\calL,\bm{s},\bm{u}),$
  where both maximums are over lattices $\calL$ with $\mathbb{E}^p \subset \calL \subset \mathbb{Z}^p$.
\end{theorem}

Unlike size properties, the separation distance of an ILDM is not inherited from that of an ILDC. Below, we provide several pivotal results concerning the separation distance of ILDMs.
Specifically, Theorem~\ref{thm:0_p:u} demonstrates that the separation distance $\rho(\calD)$ is unaffected by the choice of the vector $\bm{u}$.

\begin{theorem} \label{thm:0_p:u}
Suppose 
$\calL$ is an IL, $\caly \subset \calG$, $\bm{s} \in \mathbb{N}^p$, $\bm{u} \in \mathbb{Z}^p$, and $\calD(\calL,\bm{s},\caly,\bm{u})$ is an ILDM in \eqref{eqn:ILDM}. 
Then 
$\rho\left\{\calD(\calL,\bm{s},\caly,\bm{u})\right\}= \rho\left\{\calD(\calL,\bm{s},\caly,\bm{0}_p)\right\}.$
\end{theorem}

Combining Theorems~\ref{thm:0_p:m}-\ref{thm:0_p:u}, $\bm{u} = \bm{0}_p$ is always an optimal choice for both $\rho$ and $m$ when identifying the ILMDM with at least $n$ points.
Hence, we initially focus on solutions with $\bm{u} = \bm{0}_p$ and later allow $\bm{u}$ to vary to refine the design and achieve a size closer to $n$. This decomposition reduces the problem to two more manageable subproblems, thereby significantly lowering computational complexity without compromising design optimality.

Next, we introduce several results concerning $\caly$ and $\bm{s}$. 
Let 
\( d^*(\gamma) = \min_{i=1,\ldots,m(\gamma)-1} \{ \gamma_{(i+1)} - \gamma_{(i)} \} \)
represent the minimum distance between two adjacent levels of $\gamma$, 
and \( d^+(\gamma) = \min_{i=1,\ldots,m(\gamma)-2} \left\{ \gamma_{(i+2)} - \gamma_{(i)} \right\} \) denote the minimum distance between two non-adjacent levels of $\gamma$.


\begin{theorem} \label{thm:sep}
For any fixed $\calG \subset \prod_{k=1}^p [0, w_k]$ and $n\geq 2$, among all ILDM $\calD$ in \eqref{eqn:ILDM} with $m(\calD)\geq n$ and $\bm{u}=\bm{0}_p$, the highest separation distance $\rho(\calD)$ can be attained by a design with $s_k \ge 2$, $\caly_{k(1)}=0$, and $\caly_{k(s_k)}=w_k$ for all $1\leq k \leq p$. 
Furthermore, for such designs, 
\begin{equation}\label{eqn:rhoD}
\rho\left\{\calD(\calL,\bm{s},\caly,\bm{0}_p)\right\} = \min\left[ \min_{\bm{x} \in \calH \setminus \bm{0}_p} \left\{ \sum_{x_k=1}  d^*(\caly_k)^2 \right\}^{1/2}, \min_{s_k>2}  d^+(\caly_k)  \right].
\end{equation} 
\end{theorem}

Thoerem~\ref{thm:sep} reveals that it is sufficient to focus on designs with $s_k \ge 2$, $\caly_{k(1)}=0$, and $\caly_{k(s_k)}=w_k$ for all $1\leq k \leq p$, as at least one optimal design must satisfy these conditions. This significantly streamlines the construction process. 
Furthermore, it follows that $\rho\{\calD(\calL,\bm{s},\caly,\bm{0}_p)\} $ is related to $\caly$ only through $d^+(\caly_k)$ and $d^*(\caly_k)$ for $k=1,\ldots,p$.

\subsection{Selection of levels}\label{sec:level}

Although the preceding results constrain our attention to ILDMs of the form in
\eqref{eqn:ILDM} with $\calL$ as an IL, $\bm{s} \in \{2,3,\ldots\}^p$, $\prod_{k=1}^p \{0,w_k\} \subset \caly \subset \calG$, and $\bm{u}=\bm{0}_p$, the optimal selection of $(\calL,\bm{s},\caly)$ remains elusive due to the large number of potential configurations. 
In this subsection, we aim to narrow the search space by addressing the selection of the optimal set of levels, $\caly$, for given $(\calL,\bm{s})$. Since the design size is unaffected by $\caly$, the primary goal in choosing $\caly$ is to maximize the separation distance.

Firstly, we endeavor to determine the optimal $\caly_k$, which relies on $\calG_k$  and $s_k$, but is independent of $\calL$.
As highlighted by Theorem~\ref{thm:sep}, the ideal choice for $\caly$ should maximize both $d^*(\caly_k)$ and $d^+(\caly_k)$ for each $k$. 
If there exists a $\tilde \caly_k$ such that both $d^*(\tilde \caly_k)$ and $d^+(\tilde \caly_k)$ achieve the highest possible values among all $\caly_k \subset \calG_k$ with $m(\caly_k)=s_k$, the $\tilde \caly_k$ is certainly an optimal choice. In this case we can confidently disregard other options for $\caly_k$ without any concern for the choice of $\calL$.
Nevertheless, for some $(\calG_k,s_k)$, the highest possible $d^*$ and $d^+$ may not be simultaneously attained by a single $\caly_k$. 
In such cases, our goal is to identify a minimal sufficient set of candidate level sets, denoted as $\calF(\calG_k, s_k)$.  
For any $\gamma \subset \calG_k$ with $m(\gamma) = s_k$, there must exist a $\tilde \gamma \in \calF(\calG_k, s_k)$ such that $d^*(\tilde \gamma) \ge d^*(\gamma)$ and $d^+(\tilde \gamma) \ge d^+(\gamma)$. 
Moreover, $\calF(\calG_k, s_k)$ must consist solely of admissible choices 
$\tilde \gamma$, where no $\gamma \subset \calG_k$ with $m(\gamma)=s_k$ satisfies $d^*(\gamma)\geq d^*(\tilde \gamma)$, $d^+(\gamma)\geq d^*(\tilde \gamma)$, and $(d^*(\gamma),d^+(\gamma)) \neq (d^*(\tilde \gamma),d^+(\tilde \gamma))$. 

We identify minimal sufficient sets by examining three distinct cases. First, as shown in Theorem~\ref{thm:fcodd}, equally spaced levels are the only admissible choice for $\caly_k$ when $\calG_k = [0, w_k]$ and $s_k$ is odd.

\begin{theorem} \label{thm:fcodd}
  Suppose $\calG_k=[0,w_k]$ and $s_k \ge 2$ is odd. Let $\bar \gamma_{i} = w_k (i-1)/(s_k-1)$ for any $1\leq i 
\leq s_k$ and $\bar \gamma = (\bar \gamma_1,\ldots,\bar \gamma_{s_k})$. 
  Then $\calF([0,w_k],s_k) = \{ \bar \gamma\}$ forms a minimal sufficient set. 
\end{theorem}

Unfortunately, when $\calG_k = [0,w_k]$ and $s_k$ is even, the highest values of both $d^*$ and $d^+$ cannot be simultaneously attained. 
Assume $0 \le v_1 \le w_k/(s_k-1)$, let $v_2=(2w_k-s_k v_1)/(s_k-2)$, $\tilde \gamma_{v_1,i} = \lfloor i/2 \rfloor v_1 + (\lceil i/2\rceil-1) v_2$ for $i =1,\ldots,s_k$, 
and $\tilde \gamma_{v_1} = \{ \tilde \gamma_{v_1,1}, \ldots, \tilde \gamma_{v_1,s_k}\}$. 
Theorem~\ref{thm:fceven} provides the admissible choices of $\caly_k$ for this scenario.

\begin{theorem} \label{thm:fceven}
Suppose $\calG_k=[0,w_k]$ and $s_k > 2$ is even. 
Then  $\calF([0,w_k],s_k) = \{ \tilde \gamma_{v_1} : 0 \leq v_1 \leq w_k/(s_k-1) \}$ forms a minimal sufficient set. 
\end{theorem}

In the set, the highest possible value of $d^*$ is $w_k/(s_k-1)$, attained by $\tilde \gamma_{w_k/(s_k-1)}$, where the levels are equally spaced. 
However, $d^+(\tilde \gamma_{w_k/(s_k-1)}) = 2w_k/(s_k-1)$ is sligtly lower than the highest possible $d^+$. 
Conversely, the highest possible value of $d^+$ is $2w_k/(s_k-2)$, which is attained by $\tilde \gamma_{0}$. 
Nevertheless, $d^*(\tilde \gamma_{0}) = 0$ is very poor. 
Thus, a balance between $d^*$ and $d^+$ must be struck in practice.

\begin{algorithm}[t]
  \caption{Determine $\calF(\calG_k, s_k)$ when $\calG_k$ is a discrete set.}
  \label{alg:fksk_ordinal}
  \begin{algorithmic}[1] 
\REQUIRE $\calG_k$ and $s_k$. \\
\STATE Initialize $\tilde d^+ \leftarrow 0$ and $\calF\leftarrow\emptyset$. 
\LOOP
	\STATE Use Algorithm~\ref{alg:algdstar} to find $\tilde d^*$, which is either the maximal $d^*(\gamma)$ given that $\gamma \subset \calG_k$, $m(\gamma)=s_k$, and $d^+(\gamma) > \tilde d^+$ or zero when such $\gamma$ does not exist. 
	\STATE If $\tilde d^*=0$, \textbf{return} $\calF$. \label{alg:fksk_ordinal:dstar}
	\STATE Use Algorithm~\ref{alg:algdplus} to find the $\tilde \gamma$ that maximizes $d^+(\gamma)$ given that $\gamma \subset \calG_k$, $m(\gamma)=s_k$, and $d^*(\gamma) \geq \tilde d^*$. \label{alg:fksk_ordinal:dplus}
	\STATE Set $\calF \leftarrow \calF \cup \{ \tilde \gamma \}$.
\ENDLOOP
  \end{algorithmic}
\end{algorithm}

\begin{algorithm}[t]
  \caption{Find the maximal $d^*(\gamma)$ such that $d^+(\gamma) > \tilde d^+$ or returns 0.}
  \label{alg:algdstar}

  \begin{algorithmic}[1] 
  \REQUIRE $\calG_k$, $s_k$, and $\tilde d^+$. \\
  \STATE Initialize $\tilde d^* \leftarrow 0$. 
  \LOOP
  \STATE  Set $\gamma_1 \leftarrow 0$ and  $\gamma_2 \leftarrow \min \{l \in \calG_k : l > \tilde d^* \}$.
  \FOR{$i=3$ to $s_k$}
  \STATE 	Set $\gamma_i \leftarrow \min \{l \in \calG_k : l> \gamma_{i-1} + \tilde d^*, l > \gamma_{i-2} + \tilde d^+ \}$ is such $l$ exists and \textbf{return} $\tilde d^*$ otherwise.
  \ENDFOR
  \STATE Set $\tilde d^* \leftarrow d^*(\gamma)$.
  \ENDLOOP
\end{algorithmic}
\end{algorithm}
\begin{algorithm}[t]
  \caption{Find the $\gamma$ that maximizes $d^+(\gamma)$ given that $d^*(\gamma) \geq \tilde d^*$.}
  \label{alg:algdplus}
  \begin{algorithmic}[1] 
  \REQUIRE $\calG_k$, $s_k$, and $\tilde d^*$. \\
  \STATE Initialize $\tilde d^+ \leftarrow 0$. 
  \LOOP
  \STATE  Set $\gamma_1 \leftarrow 0$, $\gamma_2 \leftarrow \min \{l \in \calG_k : l \ge \tilde d^* \}$.
  \FOR{$i=1$ to $s_k-1$}
  \STATE 	Set $\gamma_i \leftarrow \min \{l \in \calG_k  : l \ge \gamma_{i-1} + \tilde d^*, l > \gamma_{i-2} + \tilde d^+ \}$ if such $l$ exists and \textbf{return} $\tilde \gamma$ otherwise.
  \ENDFOR
  \STATE Set $\gamma_{s_k} \leftarrow \max(\calG_k)$, $\tilde d^+ \leftarrow d^+(\gamma)$, and $\tilde \gamma \leftarrow \gamma$.
  \ENDLOOP
\end{algorithmic}
\end{algorithm} 

When $\calG_k \neq [0,w_k]$, i.e., the $k$th variable is drawn from a discrete set, the problem becomes considerably more intricate, and a straightforward characterization of admissible choices is not readily available.
While it is possible to exhaustively evaluate all potential $\gamma_k$ and identify the admissible ones, this approach is computationally inefficient, particularly when $\calG_k$ contains a large number of elements.
We propose using Algorithm~\ref{alg:fksk_ordinal} to efficiently compute $\calF(\calG_k, s_k)$, which significantly outperforms exhaustive search and remains computationally efficient for large $m(\calG_k)$. 
In this algorithm, we first identify the $\caly_k$ that yields the highest $d^*$. Subsequently, we iteratively reduce $d^*$ and identify additional admissible choices.
Algorithm~\ref{alg:fksk_ordinal} invokes two auxiliary algorithms. The first, Algorithm~\ref{alg:algdstar}, is designed to identify the maximal $d^*(\gamma)$, given that $d^+(\gamma) > \tilde d^+$ for a specified threshold $\tilde d^+$.
It is worth noting that, in some cases, no $\gamma$ may satisfy  $d^+(\gamma) > \tilde d^+$. In such instances, Algorithm~\ref{alg:algdstar} will return 0.
The second auxiliary algorithm, Algorithm~\ref{alg:algdplus}, aims to identify the $\gamma$ that maximizes $d^+(\gamma)$, given the constraint $d^*(\gamma) \geq \tilde d^*$. Importantly, whenever Algorithm~\ref{alg:algdplus} is invoked, a feasible $\gamma$ satisfying this condition is guaranteed to exist. 
An illustration of Algorithm~\ref{alg:fksk_ordinal} is provided in the appendix.
Lastly, Theorem~\ref{thm:fo} confirms that Algorithm~\ref{alg:fksk_ordinal} does produce the minimal sufficient set.

\begin{theorem} \label{thm:fo}
Suppose $\calG_k \subset [0, w_k]$ is a finite set, $2 \le s_k \le m(\calG_k)$, and $\calF(\calG_k, s_k)$ is obtained from Algorithm~\ref{alg:fksk_ordinal}. 
Then the $\calF(\calG_k, s_k)$ forms a minimal sufficient set.
\end{theorem}

\begin{algorithm}[t!]
  \caption{Find the $\caly$ 
that maximizes the $\rho\{\calD(\calL,\bm{s},\caly,\bm{0}_p)\}$ for given $\calL$ and $\bm{s}$.}
  \label{alg:Findy}
  \begin{algorithmic}[1] 
  \REQUIRE $\calG=\prod_{k} \calG_k$, $\calL$, $\bm{s}=(s_1,\ldots,s_p)$. \\
  \STATE For each ordinal dimension $k$, obtain $\calF(\calG_k,s_k)$ using Algorithm~\ref{alg:fksk_ordinal} if it has not been obtained before. 
  \STATE Initialize $\rho^{(0)} \leftarrow 0$ and $t \leftarrow 1$. 
  \STATE For $k=1,\ldots,p$, set $\caly_k\s{1}$ be the $\gamma \in \calF(\calG_k,s_k)$ that maximizes $d^*(\gamma)$.  \label{Findy:init}
\LOOP
  \STATE Set $\caly\s{t} = \prod_{k=1}^{p} \caly\s{t}_{k}$ and compute $\rho\s{t} \leftarrow \rho \{ D ( \calL,\bm{s},\caly\s{t}, \bm{0}_p ) \}$.
  \STATE If $\rho\s{t} > \rho\s{t-1}$, set $\tilde{\rho} \leftarrow \rho\s{t}$ and $\tilde\caly \leftarrow \caly\s{t}$. 
  Otherwise, \textbf{return} $\tilde{\rho}$ and $\tilde\caly$. \label{Findy:noimprovement}
  \STATE If $\rho\s{t} = \min_{\bm{x} \in \calH \setminus \bm{0}_p} \{ \sum_{x_k=1} d^*(\caly_k\s{t})^2 \}^{1/2}$, \textbf{return} $\tilde{\rho}$ and $\tilde\caly$. \label{Findy:starbound}
  \STATE If $d^+(\caly_k^{(t)}) = \rho^{(t)} = \max_{\gamma \in \calF(\calG_k,s_k)} d^+(\gamma)$ for a $k$, \textbf{return} $\tilde{\rho}$ and $\tilde\caly$.
  \STATE Set 
$\caly\s{t+1}_{k} \leftarrow \caly\s{t}_{k}$ for $k$ such that $d^+(\caly\s{t}_{k}) > \rho^{(t)}$. \label{Findy:retain}
  \STATE  For $k$ such that $\calG_k \neq [0,w_k]$ and $d^+( \caly\s{t}_k )= \rho\s{t}$, 
set $\caly\s{t+1}_{k}$ be the $\gamma \in \calF(\calG_k,s_k)$ that has the highest $d^*(\gamma)$ conditioning on that $d^+(\gamma) > d^+(\caly\s{t}_{k})$. \label{Findy:raiseo}
  \STATE Set $\calA_\text{c}^{(t)} \leftarrow \{ k : \calG_k = [0,w_k], d^+( \caly\s{t}_k )= \rho\s{t} \}$ and
$\caly\s{t+1}_{k} \leftarrow \caly\s{t}_{k}$ for $k \in \calA_\text{c}^{(t)}$.  
  \STATE If $\rho\{D(\calL,\bm{s},\caly\s{t+1},\bm{0}_p)\} < \rho\s{t}$, \textbf{return} $\tilde{\rho}$ and $\tilde\caly$.
  \IF{$\calA\s{t}_{\text{c}} \neq \emptyset$}
  \STATE  Find $\hat d^+_\text{c}$, the highest $d^+_\text{c}$ subjecting to $d^+_\text{c} \geq \rho\{D(\calL,\bm{s},\caly\s{t+1},\bm{0}_p)\}$, $d^+_\text{c} \leq \min_{k \notin \calA\s{t}_{\text{c}}} d^+(\caly\s{t+1}_k)$,  $d^+_\text{c} \leq \min_{k \in \calA\s{t}_{\text{c}} } \{ 2w_k/(s_k-2) \}$, and $(d^+_\text{c})^2 \le  \sum_{ k \in \calA\s{t}_{\text{c}}}  x_k\{w_k- d^+_\text{c}(s_k-2)/2 \}^2 +  \sum_{ k \notin \calA\s{t}_{\text{c}} }  x_k d^*(\caly_k\s{t+1})^2$ for any $\bm{x} \in \calH \backslash \bm{0}_p$. \label{eq:cons4}
    \STATE  For each $k \in \calA\s{t}_\text{c}$, set $\caly\s{t+1}_{k}$ be the $\gamma \in \calF(\calG_k,s_k)$ such that $d^+(\gamma) = \hat d^+_\text{c}$.  
  \ENDIF
  \STATE Update $t \leftarrow t+1$.
  \ENDLOOP 
\end{algorithmic}
\end{algorithm}

Thus far, for both continuous and ordinal variables, we have derived the minimal sufficient set $\calF(\calG_k, s_k)$. Next, we introduce Algorithm~\ref{alg:Findy}, which identifies the optimal $\caly_{\bm{s}} \subset \prod_{k=1}^p \calF(\calG_k,s_k)$ that maximizes the separation distance given $\calL$ and $\bm{s}$. 
As we shall demonstrate in Example~\ref{eg:f1}, the majority of optimal designs correspond to the $\caly_k$ with the maximal $d^*$ within $\calF(\calG_k, s_k)$. Therefore, the algorithm begins by selecting $\prod_{k=1}^p \caly_k$ with the highest $d^*$ for each $k$, as specified in Line~\ref{Findy:init}. We then explore other choices, starting from those with higher $d^*$ and progressing to lower values. 
For ordinal variables, we code the elements of $\calF(\calG_k,s_k)$ in ascending order. 
If the separation distance $\rho(\calD)$ is constrained by $d^*$, i.e., $\rho\s{t} = \min_{\bm{x} \in \calH \setminus \bm{0}_p} \{ \sum_{x_k=1} d^*(\caly_k\s{t})^2\}^{1/2}$, further reducing $d^*$ while increasing $d^+$ will not yield a better design. In this case, the algorithm terminates, as shown in Line~\ref{Findy:starbound}.
When this condition is not met, we retain the current $\caly_k$ for dimensions $k$ where $d^+(\caly_k) > \rho(\calD)$, as shown in Line~\ref{Findy:retain}. Simultaneously, for other dimensions $k$, we replace $\caly_k$ with a new choice that has a higher $d^+$.
For ordinal variables, we switch to the next choice,
as described in Line~\ref{Findy:raiseo}.
For even-level continuous dimensions $k$, the $d^+(\caly_k\s{t+1})$ can be switched to any value no higher than $2w_k/(s_k-2)$. 
It is reasonable to increase the $d^+(\caly_k\s{t+1})$ for all such dimensions to the same quantity, say $d^+_\text{c}$, since $\rho\s{t+1}$ is no higher then the minimum $d^+(\caly_k\s{t+1})$. 
The optimal $d^+_\text{c}$ should maximize the separation distance in (\ref{eqn:rhoD}) subject to the condition
$\rho\s{t} \le d^+_\text{c} \leq \min_{k \in \calA\s{t}_{\text{c}} } \{ 2w_k/(s_k-2) \}$. 
Therefore, we use a quadratic optimization procedure to find the optimal $d^+_\text{c}$. 
Finally, switching to a lower $d^*$ while increasing $d^+$ may not always yield a better design. 
If $\rho$ does not increase, we retain the currerently best $\caly$ and terminates the algorithm, as indicated in Line~\ref{Findy:noimprovement}.

\subsection{Selection of lattices and numbers of levels} \label{sec:alg}

In this subsection, we propose our method to select $\calL$ and $\bm{s}$ for given $p$, $n$, and $\calG$. 
Drawing from Theorems~\ref{thm:0_p:m}-\ref{thm:sep}, we first determine the optimal $D(\calL,\bm{s},\caly,\bm{0}_p)$ in \eqref{eqn:ILDM} ensuring that $s_k \ge 2$, $k=1,\ldots, p$. 
Upon obtining the optimal design, or a set of optimal designs, we proceed to identify the translation that minimizes the discrepancy between the design size and $n$. The procedure for determining the optimal $\bm{u}$ is outlined in Algorithm~\ref{alg:chooseu}.

\begin{algorithm}[t]
  \caption{Find the design whose size is the closest to $n$.}
  \label{alg:chooseu}
  %
\begin{algorithmic}[1] 
\REQUIRE $p$, $n$, $t$, and $\calL^{(i)}$, $\bm{s}^{(i)}$, and $\caly^{(i)}$ for  $i=1,\ldots,t$. 
\FOR {$i=1$ to $t$} \label{alg1:ustart}
 \STATE Set $\bm{u}\s{i}$ be the $\bm{u} \in \{0,1\}^p$ that minimizes $m(\calL^{(i)},\bm{s}^{(i)},\bm{u})$ while satisfying $m(\calL^{(i)},\bm{s}^{(i)},\bm{u}) \ge n$ and set $m\s{i} \leftarrow m(\calL^{(i)},\bm{s}^{(i)},\bm{u}\s{i})$. 
\ENDFOR
\STATE Find $\hat{i}$, the $1\leq i\leq t$ that minimizes $m\s{i}$.
\STATE \textbf{Return} $\calL^{(\hat{i})}$, $\bm{s}^{(\hat{i})}$, $\caly^{(\hat{i})}$, $\bm{u}^{(\hat{i})}$, and $m\s{\hat{i}}$. 
\end{algorithmic}
\end{algorithm} 

We outline three primary algorithms for determining the optimal $\calL$ and $\bm{s}$. 
First, when $2\leq p\leq 5$, we propose to exhaustively search all ILs. 
The procedure for this search is presented in Algorithm~\ref{alg:pless5}.
For each candidate $\calL$, we suggest evaluating the $\bm{s}$ in increasing lexicographical order. 
During the optimization process, we maintain $\hat{\rho}$, which denotes the best separation distance achieved among the designs explored up to that point.
Based on Theorem~\ref{thm:sep}, 
$\rho\left\{\calD(\calL,\bm{s},\caly,\bm{0}_p)\right\} \leq d^+(\caly_k) $ for any $k$. 
Therefore, the algorithm restricts the search to $\bm{s}$ such that, for each $k$, the highest possible $d^+$ from the choices of $\calF(\calG_k,s_k)$ is at least $\hat\rho$. For continuous variables, this implies $s_k \leq \lfloor 2w_k/\hat{\rho} + 2\rfloor$.

\begin{algorithm}[t]
  \caption{Construct ILMDM for $2\le p \le 5$.}
  \label{alg:pless5}
  %
  \begin{algorithmic}[1] 
  \REQUIRE $p$, $n$, $\calG$, $\calF$, and the list of all $p$-dimensional ILs $\{\calL_i\}_{i=1}^N$.
\STATE Initialize $\hat{\rho}=0$ and ${\cal S}_k = \{2,\ldots,m(\calG_k)\}$ for all $k$.  \label{alg6:initial}
\FOR {$i=1$ to $N$}   
 \STATE Initialize $\tilde{\bm{s}} \leftarrow \bm{0}_p$. 
 \WHILE{ $\{\bm{z} \in \cals: m(\calL_i,\bm{z},\bm{0}_p) \ge n, \bm{z} > \bm{\tilde s}\} \neq \emptyset$ }
  \STATE Set $\bm{s} \leftarrow \min \{\bm{z} \in \cals: m(\calL_i,\bm{z},\bm{u}) \ge n, \bm{z} > \bm{\tilde s}\}$. 
  \STATE Run Algorithm~\ref{alg:Findy} to obtain $\tilde\caly$, the $\caly$ that maximizes $\rho\{\calD(\calL_i,  \bm{s}, \allowbreak \caly, \bm{0}_p)\}$. 
  \IF{ $\rho\{\calD(\calL_i,\bm{s},\caly,\bm{0}_p)\} > \hat{\rho} $} \label{alg6:checkrho}
  \STATE Update $\hat{\rho} \leftarrow \rho\{\calD(\calL_i,\bm{s},\caly,\bm{0}_p)\}$ and set $t \leftarrow 1$, $\hat \calL^{(1)} \leftarrow \calL_i$, $\bm{\hat s}^{(1)} \leftarrow \bm{s}$, $\hat \caly^{(1)} \leftarrow \tilde\caly$. 
  \STATE For each $k$ such that $2<m(\calG_k)<\infty$, set ${\cal S}_k \leftarrow {\cal S}_k \cap \{ z \in \mathbb{N} : \max_{\gamma \in \calF(\calG_k,z)} d^+(\gamma) \ge \hat{\rho} \}$. 
  \STATE For each $k$ such that $m(\calG_k)=\infty$, set ${\cal S}_k \leftarrow {\cal S}_k \cap [ 0, \lfloor 2 \max(\calG_k)/\hat{\rho} \allowbreak  + 2\rfloor ]$. \label{alg6:updateS}
  \ELSIF {$\rho\{\calD(\calL_i,\bm{s},\caly,\bm{0}_p)\} = \hat{\rho}$ }
  \STATE Set $t \leftarrow t+1$, $\hat\calL\s{t} \leftarrow \calL_i$, $\bm{\hat s}\s{t} \leftarrow \bm{s}$, and $\hat \caly\s{t} \leftarrow \tilde\caly$.
  \ENDIF
  \STATE Set $\bm{\tilde s} \leftarrow \bm s$ and then set $\tilde s_k \leftarrow +\infty$.
 \ENDWHILE
\ENDFOR \label{alg6:endfor}
\STATE Run Algorithm~\ref{alg:chooseu} to obtain $\hat{\calL}$, $\hat{\bm{s}}$, $\hat\caly$, $\hat{\bm{u}}$, and $\hat{m}$. \label{alg6:findbest}
\STATE \textbf{Return} $\calD(\hat{\calL},\hat{\bm{s}}, \hat\caly, \hat{\bm{u}})$, $\hat{\rho}$, and $\hat{m}$.
\end{algorithmic}
\end{algorithm} 

The number of distinct ILs increases exponentially with the dimensionality $p$, resulting in a rapid escalation of the computational time required to execute Algorithm~\ref{alg:pless5}. 
Consequently, we recommend employing Algorithm~\ref{alg:pless5} exclusively for dimensions  $2 \le p \le 5$. 
In contrast, the computational burden associated with increasing the number of points $n$ is relatively minimal, allowing Algorithm~\ref{alg:pless5} to be effectively applied to large-$n$ problems.

We present three examples to illustrate the application of Algorithm~\ref{alg:pless5}, one of them is shown in the appendix. 

\begin{example} \label{eg:f1}
Running Algorithm~\ref{alg:pless5} for all cases with $2\leq p\leq 5$, $4\leq n \leq 1000$, and $\calG = \{0,0.01,0.02,\ldots,1\} \times [0,1]^{p-1}$,   
we observe that among the choices of $\calF(\calG,\bm{s})$, the selection that attains the highest possible $d^*$ is chosen significantly more often than others. 
Specifically, for $p=2,3,4,5$, out of the 997 designs there are only $134,33,66,16$ designs, respectively, where at least one $\caly_k$ does not attain the highest possible $d^*$.   
\end{example}

\begin{example} \label{eg:notf1}
Suppose $p = 3$, $n = 10$, and $\calG = [0,1]^3$. The ILMDC obtained from \citet{he2019interleaved} is
$\calD\{ \calL(\bm{G}_{3,2}), (2,3,3), \{0, 1\} \times \{0,0.5, 1\}^2, (0,0,0)\}$,
which has $\rho = 0.7071$, and $m = 10$.
In contrast, the ILMDM found from Algorithm~\ref{alg:pless5} is $\calD\{ \calL(\bm{G}_{3,3}), (4,3,3), \{0,0.25,0.75, 1\} \times \{0,0.5, 1\}^2, \allowbreak (0,0,0)\}$, which has $\rho=0.75$, and $m=10$. 
Both $\bm{G}_{3,2}$ and $\bm{G}_{3,3}$ are given in \eqref{eqn:G:3}. 
\end{example}

Example~\ref{eg:notf1} implies that Algorithm~\ref{alg:pless5} can sometimes generate designs that outperform ILMDC in terms of separation distance. 
This result can be attributed to the fact that ILMDC requires the levels to be equally spaced, while ILMDM relaxes this constraint.
However, 
Algorithm~\ref{alg:pless5} incurs slightly higher computational costs due to the larger number of configurations considered.

Next, we present an algorithm for $6\leq p\leq 8$. Given that Algorithm~\ref{alg:pless5} becomes computationally intensive for higher-dimensional cases, we propose two strategies to enhance its efficiency.
First, we observe that in executing Algorithm~\ref{alg:pless5}, the larger $\hat\rho$ we have obtained, the more choices of $\bm{s}$ can be eliminated. 
Therefore, early identification of a near-optimal design allows for the pruning of less promising configurations. To expedite this process, we propose initially testing three simple yet promissing types of ILs.
Let 
\begin{equation}\label{eqn:G:3}
 \bm{G}_{p,1} = \bm{I}_p, 
\quad \bm{G}_{p,2} = \begin{pmatrix}
    \bm{I}_{p-1}& \bm{1}_{(p-1) \times 1}\\
    \bm{0}_{1\times (p-1)}&2
    \end{pmatrix}, \quad
 \bm{G}_{p,3} = \begin{pmatrix}
      1&\bm{1}_{1 \times (p-1)}\\
      \bm{0}_{(p-1)\times 1}&2\bm{I}_{p-1}
      \end{pmatrix}. 
\end{equation}
The lattice $\calL(\bm{G}_{p,1}) = \mathbb{Z}^p$ represents the full grid. The lattice $\calL(\bm{G}_{p,2})$ consists of half of the points $\bm{x}$ from $\calL(\bm{G}_{p,1})$, specifically those for which the sum  $\sum_{k=1}^p x_k$ is even. The lattice $\calL(\bm{G}_{p,3}) = \mathbb{E}^p \cup ( \mathbb{E}^p \oplus \bm{1}_p )$.
We propose using these three lattices, $\calL(\bm{G}_{p,1})$, $\calL(\bm{G}_{p,2})$, and $\calL(\bm{G}_{p,3})$, as the initial candidates in our search process.

\begin{algorithm}[t!]
  \caption{Check for the existence and find the best $\calL$ for the $(q,r,\bm{s})$ choice.}
  \label{alg:computeL}
  %
  \begin{algorithmic}[1] 
  \REQUIRE $p$, $n$, $\caly$, $q$, $r$, $\bm{s}$, $\hat\rho$. 
\STATE 
Initialize $\calH \leftarrow \{ (x_1,\ldots,x_p)  : 0 \leq x_k \leq 1 \text{ if } d^*(\caly_k)  \text{  is among the top } \allowbreak \text{   $r$ values of }$,  $d^*(\caly_1),\ldots, 
d^*(\caly_p) \text{ and } \allowbreak x_k=0 \text{ otherwise for } k=1,\ldots,p \}$ and
$\bar{\calH} \leftarrow \{ \bm{e}_k : 1\leq k\leq p \} \setminus \calH$.   
\WHILE{$m(\calH)<2^q$} 
  \STATE If there is an $\bm{x} \in \calH$ such that $\sum_{k=1}^p \{ d^*(\caly_k) x_k \}^2 < \hat\rho^2$,  \textbf{terminate} the algorithm by claiming that eligible $\calL$ does not exist. 
  \STATE Set $\bm{\bar x}$ be the vector $\bm{x}$ that has the lowest $\sum_{k=1}^p \{ d^*(\caly_k) x_k \}^2$ among $\{0,1\}^p \setminus \calH \setminus \bar{\calH}$. 
  \STATE Set $E \leftarrow 1$, $\calH_0 \leftarrow \calH$, and $\calH_j \leftarrow \emptyset$ for $j = 1,\ldots,2^{p-q}-1$. 
  \FOR{$\bm{x} \in \bar{\calH} \cup \{\bm{\bar x}\}$}  
    \STATE Set $j\leftarrow 1$.
    \WHILE{$j<2^{p-q}$} \label{step:computeL:begin}
      \STATE If $\calH_j = \emptyset$, set $\calH_j \leftarrow \{\bm{x}\}$, \textbf{break} the loop on $j$ and continue to the next $\bm{x}$.
      \STATE Set  $\tilde \calH_j \leftarrow [ \cup_{\bm{z}\in\mathbb{Z}^p}  \{ \calH_j \oplus \bm{ x} \oplus (2\bm{z}) \} ] \cap \{0,1\}^p$, $\hat \calH \leftarrow [ \cup_{\bm{z}\in\mathbb{Z}^p} \cup_{\tilde{\bm{x}} \in \tilde \calH_j} \{ \calH \oplus \tilde{\bm x} \oplus (2\bm{z}) \} ] \cap \{0,1\}^p$, $\hat \calH_0 \leftarrow \calH_0 \cup \hat\calH$,  and $\hat \calH_k \leftarrow [ \cup_{\bm{z}\in\mathbb{Z}^p} \cup_{\hat{\bm{x}} \in \hat\calH} \{ \calH_k \oplus \bm{\hat x} \oplus (2\bm{z}) \} ] \cap \{0,1\}^p$ for $k=1,\ldots,2^{p-q}-1$.  
      \STATE If the $\hat \calH_0,\hat\calH_1,\ldots,\hat\calH_{2^{p-q}-1}$ are non-overlapping, \textbf{break} the loop on $j$.
      \STATE Set $j \leftarrow j+1$.
      \ENDWHILE \label{step:computeL:end}
    \STATE If $j=2^{p-q}$, set  $\tilde \calH \leftarrow [ \cup_{\bm{z}\in\mathbb{Z}^p}  \{ \calH \oplus \bm{ x} \oplus (2\bm{z}) \} ] \cap \{0,1\}^p$, $\calH \leftarrow \calH \cup \tilde \calH$, $E \leftarrow 0$, and \textbf{break} the loop on $\bm{x}$. 
    \STATE Set $\calH_k \leftarrow \hat\calH_k$ for $k=0,\ldots,2^{p-q}-1$. 
  \ENDFOR
  \STATE If $E=1$, update $\bar{\calH} \leftarrow \bar{\calH} \cup \{ \bar{\bm{x}}\}$. 
\ENDWHILE
\STATE If there is an $\bm{x} \in \calH$ such that $\sum_{k=1}^p \{ d^*(\caly_k) x_k \}^2 < \hat\rho^2$,  \textbf{terminate} the algorithm by claiming that eligible $\calL$ does not exist. 
\STATE \textbf{Return} the $\calL \leftarrow \sum_{\bm{z} \in \mathbb{Z}^p} \{ \calH \oplus (2\bm{z}) \}$. 
\end{algorithmic}
\end{algorithm}

As a secondary strategy to accelerate the search, we propose concentrating on designs derived from a select subset of promising IL types.
Specifically, we evaluate all viable $(q, r, \bm{s})$ configurations, giving priority to the IL that maximizes separation distance in each iteration.
Algorithm~\ref{alg:computeL} details the procedure for identifying the appropriate IL. This involves systematically assigning elements from $\{0,1\}^p$  to either the set $\calH = \calL \cap \{0,1\}^p$ or its complement, $\bar{\calH} = \{0,1\}^p \setminus \calH$.
Note that the lattice $\mathbb{Z}^p$ can be partitioned into $2^{p-q}$ subsets $\calL_0, \ldots, \calL_{2^{p-q}-1}$, where $\calL_0 = \calL$ and each $\calL_j$ for $j > 0$ can be expressed as $\calL_j = \calL \oplus \bm{x}$, with $\bm{x} \in \{0,1\}^p$~\citep{he2019interleaved}.
Since $\calL$ constitutes a lattice and each $\calL_j$ is a translation of $\calL$, it follows that if $\bm{x} \in \calL_0$ and $\bm{z} \in \calL_j$, then $\bm{x} + \bm{z} \in \calL_j$.
Additionally, if $\bm{x}$ and $\bm{z}$ both belong to $\calL_j$ for a $j$, then $\bm{x} + \bm{z} \in \calL_0$. Consequently, these lattice properties restrict how elements from $\{0,1\}^p$ are distributed between $\calH$ and $\bar{\calH}$.

Initially, $\calH$ must incorporate the origin vector $\bm{0}_p$. Considering that $r(\calL) = r$, the initial $r$ unit vectors $\bm{e}_k$, which maximize $d^*(\caly_k)$, are allocated to $\calH$. The remaining $p - r$ unit vectors are consequently assigned to $\bar{\calH}$.
Subsequently, we iteratively assign the vector $\bm{x}$, which minimizes $\sum_{k=1}^p \{ d^*(\caly_k) x_k \}^2$, from the residual elements in $\{0,1\}^p$. To optimize $\rho(\calL)$, preference is given to allocating $\bm{x}$ to $\bar{\calH}$ when possible.
During the allocation of each vector $\bm{x}$, Lines~\ref{step:computeL:begin}-\ref{step:computeL:end} facilitate a meticulous search for a suitable subset $\calH_j = \calL_j\cap \{0,1\}^p$, aiming to incorporate $\bm{x}$ while adhering to the lattice structure and previously established rules.
Algorithm~\ref{alg:computeL} terminates once $\calH$ encompasses $2^q$ elements, as $q(\calL) = q$. Nevertheless, should any vector $\bm{x} \neq \bm{0}_p$ be allocated to $\calH$ such that $\sum_{k=1}^p \{ d^*(\caly_k) x_k \}^2 < \hat\rho^2$, the algorithm prematurely halts, indicating the absence of a viable $\calL$.
Despite the allocation of an element $\bm{x}$ with low $\sum_{k=1}^p \{ d^*(\caly_k) x_k \}^2$ to $\calH$, the algorithm persists in attempting to allocate the smallest remaining elements to $\bar{\calH}$. This strategic assignment guarantees that the resulting $\calL$ not only maximizes the separation distance but also enhances the second-lowest pairwise distance, thereby substantially elevating the design's overall quality.

\begin{algorithm}[t!]
  \caption{Find the otpimal ILDMs for several types of simple choices.}
  \label{alg:initialRho}
  \begin{algorithmic}[1] 
  \REQUIRE $p$, $n$, $\calG$. 
  \STATE Initialize the set of condidate results $\cal C = \emptyset$.
\FOR {$i=1$ to $3$} \label{step:initial:for}
  \STATE Set $\bm{s} \leftarrow \bm{2}_p$. 
  \WHILE {$m\{\calL(\bm{G}_{p,i}),\bm{s},\bm{0}_p\}<n$ and $s_k < m(\calG_k)$ for at least one $k$} 
  \STATE For $k=1,\ldots,p$, set $\caly_k$ be the one with the highest $d^*$ among those in $\calF(\calG_k,s_k)$.
  \STATE Set $s_k \leftarrow s_k+1$ for every $k$ such that $d^*(\caly_k) = \max_{j=1}^p d^*(\caly_j)$. 
  \ENDWHILE
  \STATE Run Algorithm~\ref{alg:Findy} to obtain $\tilde \caly$, the $\caly$ that maximizes $\rho[\calD\{\calL(\bm{G}_{p,i}),  \bm{s},\allowbreak \caly,\bm{0}_p\}]$.
   \STATE Add $( \rho[\calD\{\calL(\bm{G}_{p,i}),\bm{s},\caly,\bm{0}_p\}], \calL(\bm{G}_{p,i}), \bm{s}, \tilde\caly )$ into the set $\cal C$. 
\ENDFOR \label{step:initial:endfor}
  \STATE Return $\cal C$. 
\end{algorithmic}
\end{algorithm}

\begin{algorithm}[t!]
  \caption{Construct the ILMDM for $6 \leq p \le 8$.}
  \label{alg:pgreater5}
  \begin{algorithmic}[1] 
\REQUIRE $p$, $n$, $\calG$, $\calF$, and $T$. 
\STATE Initialize  ${\cal S}_k \leftarrow \{2,\ldots,m(\calG_k)\}$ for all $k$. \label{alg8:start}
\STATE Determine the set of condidate results $\cal C$ by Algorithm~\ref{alg:initialRho}. 
   \STATE Set $\hat\rho$ be the smallest $\rho$ if $\cal C$ contains fewer than $T$ elements. Otherwise  set $\hat\rho$ be the $T$th highest $\rho$ of the elements in $\cal C$, and remove the elements whose $\rho$ is lower than $\hat\rho$. \label{alg8:upCrho}
   \STATE For each $k$ with $2<m(\calG_k)<\infty$, set ${\cal S}_k \leftarrow {\cal S}_k \cap \{ z \in \mathbb{N} : \max_{\gamma \in \calF(\calG_k,z)} d^+(\gamma) \ge \hat{\rho} \}$.  \label{alg8:upS1}  
   \STATE For each $k$ with $m(\calG_k)=\infty$, set ${\cal S}_k \leftarrow {\cal S}_k \cap [ 0, \lfloor 2\max(\calG_k)/\hat{\rho} + 2\rfloor ]$.  \label{alg8:upS2}
\FOR {$q=(p-1)$ to $1$}   
 \STATE Initialize $\bm{s} \leftarrow \bm{2}_p$. 
 \WHILE{$\{ \bm{z} \in \cals: \bm{z} > \bm{s},  z_p \geq 2 \lceil2^{p-q-1}n \{ \prod_{k=1,k\neq j}^{p-1}(2 \lceil z_k/2 \rceil) \}^{-1} \rceil -1 \} \neq \emptyset$} \label{step:pgreater5:while}
\STATE   Set $\bm{s} \leftarrow \min \{ \bm{z} \in \cals: \bm{z} > \bm{s},  z_p \geq 2 \lceil2^{p-q-1}n  \{ \prod_{k=1, k \neq j}^{p-1}(2  \allowbreak \lceil z_k/2 \rceil)  \}^{-1} \rceil -1 \} $, and set $E \leftarrow 0$.  \label{step:pgreater5:s}
\STATE  For each $k$ such that $m(\calG_k)<\infty$, set $\caly_k$ be the one in $\calF(\calG_k,s_k)$ that has the largest $d^*$ provided that $d^+ \geq \hat\rho$.
\STATE  For each $k$ such that $m(\calG_k)=\infty$, set $\caly_k$ be the one in $\calF(\calG_k,s_k)$ such that $d^* = w_k - \hat{\rho}\{\max({\cal S}_k)-2\}/2$. 
  \FOR {$r$ from $q-1$ to $0$}
   \STATE Use Algorithm~\ref{alg:computeL} to find the best $\calL$ for the $(q,r,\bm{s})$ choice if such a lattic exists. 
   \STATE If such $\calL$ does not exist, \textbf{continue} to the next choice of $r$. 
   \STATE Set $E \leftarrow 1$ and compute $m(\calL,\bm{s},\bm{0}_p)$. 
   If $m(\calL,\bm{s},\bm{0}_p)<n$, \textbf{goto} Line~\ref{step:pgreater5:while}. \label{step:pgreater5:goto}
   \STATE Run Algorithm~\ref{alg:Findy} to obtain $\tilde\caly$, the $\caly$ that maximizes $\rho = \rho\{\calD(\calL,\bm{s},\caly, \bm{0}_p)\}$. 
   \STATE Add $( \rho, \calL, \bm{s}, \tilde\caly ) \ge \hat \rho$ into the set $\cal C$. 
   \STATE Update $\hat{\rho}$, $\cal C$, and ${\cal S}_1,\ldots,{\cal S}_p$ according to Lines~\ref{alg8:upCrho}-\ref{alg8:upS2}.
     \ENDFOR
  \STATE If $E = 0$, set $s_p \leftarrow \infty$. 
 \ENDWHILE
\ENDFOR \label{alg8:restore}
\STATE Submit elements of $\cal C$ to Algorithm~\ref{alg:chooseu} to obtain $\hat{\calL}$, $\hat{\bm{s}}$, $\hat\caly$, $\hat{\bm{u}}$, and $\hat{m}$.  
\STATE \textbf{Return} $\calD(\hat{\calL},\hat{\bm{s}}, \hat\caly, \hat{\bm{u}})$, $\hat{\rho}$, and $\hat{m}$.
\end{algorithmic}
\end{algorithm}

Algorithm~\ref{alg:pgreater5} outlines our second primary algorithm for constructing ILMDM, which exploits the two strategies. 
We recommend setting $T=1$ to focus solely on designs with the highest separation distance. 
After fully exploiting the potentially optimal choices with $r = q = p$ 
from the options with $\calL(\bm{G}_{p,1})$ 
in 
Algorithm~\ref{alg:initialRho}, we restrict our focus to $q<p$ cases in the subsequent steps.
In Line~\ref{step:pgreater5:s}, we impose sample size constraints from Theorems~\ref{thm:m:s} to refine the selection of $(q, r, \bm{s})$. Furthermore, in Line~\ref{step:pgreater5:goto}, we terminate the search for smaller values of $r$, as Theorem~\ref{thm:m:r} indicates that reducing $r$ tends to decrease the design size.

Algorithm~\ref{alg:pgreater5} does not guarantee to detect the ILDM with the highest separation distance, as the optimal $\calL$ found by the algorithm is only optimal when paired with the $\caly$ that maximizes $d^*$.
Nevertheless, we believe that in most cases, Algorithm~\ref{alg:pgreater5} can reliably identify the ILMDM. 
For instance, comparisons with designs obtained from Algorithm~\ref{alg:pless5} demonstrate that Algorithm~\ref{alg:pgreater5} successfully identifies the ILMDM across all scenarios with $2 \le p \le 5$, $2 \le n \le 1000$, and $\calG = [0,1]^p$ or $\calG = \{0,0.1,0.2,\ldots,1\}^p$. 
This is presumably due to the fact that most optimal designs correspond to the $\caly$ that maximizes $d^*$, as demonstrated in Example~\ref{eg:f1}. 
Unfortunately, while Algorithm~\ref{alg:pgreater5} is computationally efficient for $p \leq 8$, its performance deteriorates significantly as $p$ increases. Therefore, we recommend using Algorithm~\ref{alg:pgreater5} primarily for cases where $6 \leq p \leq 8$.

\begin{algorithm}[t!]
  \caption{Construct the ILMDM for $p>8$.}
  \label{alg:pgreater8}
  %
  \begin{algorithmic}[1] 
  \REQUIRE $p$, $n$, $\calG$, $T$. 
  \STATE Permute the dimensions of $\bm{w}$ and $\calG$ so that $w_1 \ge \ldots \ge w_p$. \label{alg3:permute}
  \STATE Run Lines~\ref{alg8:start}-\ref{alg8:restore} of Algorithm~\ref{alg:pgreater5} to obtain a list of at least $T$ ILMDMs for $p=8$, $n=n$, and $\calG = \prod_{k=1}^8 \calG_k$ 
  and name them $(\calL^{(i,8)}, \bm{s}^{(i,8)}, \caly^{(i,8)})$, $i=1, \ldots, \hat T$. 
  \label{alg3:runalg2}
   \STATE Initialize $\hat{\rho}=0$.
  \FOR {$i=1$ to $\hat T$} 
  \FOR {$j=8$ to $p-1$} 
\STATE  Set $\calx \leftarrow ( \calL^{(i,j)} \cap \{0,1\}^j ) \setminus \bm{0}_j $, 
$\calL' \leftarrow \{ \bm{0}_j \}$, and $\calL'' \leftarrow \emptyset$. \label{alg3:par:start}
\WHILE{ $\calx \neq \emptyset$ }
 \STATE Find the $\bm{x} \in \calx$ that has the lowest $\sum_{x_k=1} d^*(\caly^{(i,j)}_k)^2$.  
 \STATE Set $\tilde{\calx} \leftarrow \calL'' \oplus \bm{x}$, $\calL' \leftarrow \calL' \cup \tilde{\calx}$, $\calL' \leftarrow \calL' \cup \{ \tilde{\bm{x}} : \tilde{\bm{x}}  = \bar{\bm{x}} + \dot{\bm{x}}, \bar{\bm{x}}, \dot{\bm{x}} \in \calL' \}$, 
$\calL'' \leftarrow \calL'' \cup \{\bm{x}\} $, $\calL'' \leftarrow \calL'' \cup \{ \tilde{\bm{x}} : \tilde{\bm{x}} \leftarrow \bar{\bm{x}} + \dot{\bm{x}}, \bar{\bm{x}} \in \calL', \dot{\bm{x}} \in \calL'' \}$, 
and $\calx \leftarrow \calx \setminus \calL' \setminus \calL''$. 
\ENDWHILE
  \STATE Set $\calL^{(i,j+1)} \leftarrow \{(\bm{y},2z) : \bm{y} \in \calL', z\in \mathbb{Z} \} \cup \{ (\bm{y},2z+1): \bm{y} \in \calL'',z \in \mathbb{Z} \}$, $\bm{s}^{(i,j+1)} \leftarrow (\bm{s}^{(i,j)},2)$, and $\caly^{(i,j+1)} \leftarrow \caly^{(i,j)} \times \{0,w_{j+1}\}$. \label{alg3:par:end}
  \ENDFOR
  \STATE Run Algorithm~\ref{alg:Findy} to obtain $\tilde{\caly}$, the $\caly$ that maximizes $\rho\{\calD(\calL^{(i,p)},\bm{s}^{(i,p)},\caly, \bm{0}_p)\}$. 
   \IF{ $\rho\{\calD(\calL^{(i,p)},\bm{s}^{(i,p)},\tilde{\caly}, \bm{0}_p)\} > \hat{\rho} $}
    \STATE Update $\hat{\rho} \leftarrow \rho\{\calD(\calL^{(i,p)},\bm{s}^{(i,p)},\tilde{\caly},\bm{0}_p)\}$, $t \leftarrow 1$, $\hat \calL^{(1)} \leftarrow \calL^{(i,p)}$, $\bm{\hat s}^{(1)} \leftarrow \bm{s}^{(i,p)}$, $\hat \caly^{(1)} \leftarrow \tilde{\caly}$. 
   \ELSIF {$\rho\{\calD(\calL^{(i,p)},\bm{s}^{(i,p)},\tilde{\caly}, \bm{0}_p)\} = \hat{\rho}$ }
    \STATE Set $t \leftarrow t+1$, $\hat\calL\s{t} \leftarrow \calL^{(i,p)}$, $\bm{\hat s}\s{t} \leftarrow \bm{s}^{(i,p)}$, and $\hat \caly\s{t} \leftarrow \tilde{\caly}$.
   \ENDIF
  \ENDFOR
\STATE Run Algorithm~\ref{alg:chooseu} to obtain $\hat{\calL}$, $\hat{\bm{s}}$, $\hat\caly$, $\hat{\bm{u}}$, and $\hat{m}$. 
\STATE Use the inverse of the permutation in Line~\ref{alg3:permute} to permute the dimensions of $\hat{\calL}$, $\hat{\bm{s}}$, $\hat\caly$, $\hat{\bm{u}}$. 
\STATE \textbf{Return} $\calD(\hat{\calL},\hat{\bm{s}}, \hat\caly, \hat{\bm{u}})$, $\hat{\rho}$, and $\hat{m}$.
\end{algorithmic}
\end{algorithm}

Finally, we present our third main algorithm, which offers improved computational efficiency compared to Algorithm~\ref{alg:pgreater5}. Based on empirical observations, for $p > 8$, unless $n$ is exceptionally large and equal weights are applied, the optimal design generally involves only two distinct values, namely $0$ and $w_k$, for the least significant variables starting from the ninth dimension onward.
Thus, our primary strategy involves augmenting the design by incorporating $0$ and $w_k$ values, derived from eight-dimensional designs. The steps are outlined in Algorithm~\ref{alg:pgreater8}.
In this algorithm, we first sort the variables by $w_k$ and obtain top eight-dimensional designs. 
We recommend setting $T=20$ such that at least 20 top eight-dimensional designs are selected. 
We then supplement the remaining dimensions sequentially, following the descending order of $w_k$. 
For each new dimension, we greedily select the design that maximizes the separation distance. 
Finally, we select the design with the highest separation distance and the closest size.

In Lines~\ref{alg3:par:start}-\ref{alg3:par:end} of Algorithm~\ref{alg:pgreater8}, the lattice $\calL^{(i,j)}$ is partitioned into two sublattices: $\calL'$, a sublattice of $\mathbb{Z}^j$ containing $\mathbb{E}^j$, and $\calL''$, a translation of $\calL'$, i.e., $\calL'' = \calL' \oplus \bm{x}$ for some $\bm{x} \in \mathbb{Z}^j$.
We then let
$\calL^{(i,j+1)} = \{(\bm{y},2z) : \bm{y} \in \calL', z\in \mathbb{Z} \} \cup \{ (\bm{y},2z+1): \bm{y} \in \calL'',z \in \mathbb{Z} \}$, 
$\bm{s}^{(i,j+1)} = (\bm{s}^{(i,j)},2)$, and $\caly^{(i,j+1)} = \caly^{(i,j)} \times \{0,w_{j+1}\}$.
Clearly, such $\calL^{(i,j+1)}$ is an IL and
$ \rho\left\{D(\calL^{(i,j+1)},\bm{s}^{(i,j+1)},\caly^{(i,j+1)},\bm{0}_{j+1})\right\} = \min \left[ \rho(\calL'), \left\{ \rho(\calL'')^2 + w_{j+1}^2 \right\}^{1/2} \right]$.
To maximize $\rho\{D(\calL^{(i,j+1)},\bm{s}^{(i,j+1)},\caly^{(i,j+1)},\bm{0}_{j+1})\}$, the optimal strategy is to assign to $\calL''$ the element $\bm{x} \in  \{ \calL^{(i,j)} \cap \{0,1\}^j \} \setminus \bm{0}_j$ that minimizes $\sum_{x_k=1} d^*(\caly^{(i,j)}_k)^2$.
The elements $\bm{x} \in \{ \calL^{(i,j)} \cap \{0,1\}^j \} \setminus \bm{0}_j$ are sorted in ascending order of $\sum_{x_k=1} d^*(\caly^{(i,j)}_k)^2$ and then assigned iteratively to either $\calL'$ or $\calL''$, with each element being placed in $\calL''$ unless its allocation to $\calL'$ is required.
Note that since $\calL'$ is a lattice and $\calL''$ is a translation of $\calL'$, for any two vectors  $\bar{\bm{x}}, \dot{\bm{x}} \in \calL'$, their sum $\bar{\bm{x}} + \dot{\bm{x}}$ must also belong to $\calL'$. 
Similarly, for any two vectors $\bar{\bm{x}}, \dot{\bm{x}} \in \calL''$, their sum $\bar{\bm{x}} + \dot{\bm{x}}$ must lie in $\calL'$. Furthermore, for any vector $\bar{\bm{x}} \in \calL'$ and $\dot{\bm{x}} \in \calL''$, their sum $\bar{\bm{x}} + \dot{\bm{x}}$ must lie in $\calL''$.
Using these three rules, after assigning an element $\bm{x}$ to $\calL''$, 
the allocation of many other elements from $\{ \calL^{(i,j)} \cap \{0,1\}^k \} \setminus \bm{0}_k$ are mandatory.  
By Theorem~\ref{thm:sep}, the design $D(\calL^{(i,j+1)},\bm{s}^{(i,j+1)},\caly^{(i,j+1)},\bm{0}_{j+1})$, constructed in this manner, attains the maximum separation distance among all designs derived from supplementing $D(\calL^{(i,k)}, \bm{s}^{(i,k)}, \caly^{(i,k)}, \bm{0}_k)$ with $0$ and $w_{j+1}$.

Compared to the ILMDC construction algorithm in \cite{he2019interleaved}, our three main algorithms introduce additional complexity for three key reasons. First, we optimize the selection of levels for each variable, while ILMDC relies on fixed, equally spaced levels. Second, our method imposes a constraint on the maximum number of levels per ordinal variable, whereas ILMDC does not. Third, we include a final step to select the design whose sample size is closest to the target, a step not considered in \cite{he2019interleaved}.

\begin{table}[t!] 
  \centering
  \caption{Nine types of ordinal variables.}     \label{tab:QQ}\par
 \begin{tabular}{|cc|} \hline 
    Name & Levels \\ \hline
    $Q_1$ & $\{0,1\}$  \\ \hline
    $Q_2$ & $\{0,0.5,1\}$  \\ \hline
    $Q_3$ & $\{0,0.1,0.3,0.6,1\}$  \\ \hline
    $Q_4$ & $\{0,0.25,0.5,0.75,1\}$  \\ \hline
    $Q_5$ & $\{0,0.2,0.3,0.5,0.7,1\}$  \\ \hline
    $Q_6$ & $\{0,0.2,0.4,0.6,0.8,1\}$  \\ \hline
    $Q_7$ & $\{0,0.1,0.3,0.5,0.7,0.9,1\}$  \\ \hline
    $Q_8$ & $\{0,0.1,0.2,0.3,0.4,0.5,0.6,0.7,0.8,0.9,1\}$  \\ \hline
    $Q_9$ & $\{0,0.02,0.03,0.05,…,0.83,0.89,0.97,1\}$  \\ \hline
  \end{tabular}
  \end{table}

  \begin{figure}[t!]
    \centering
    \begin{minipage}{1\linewidth}
      \centering	
      \subfigure[$p=3$, $\calG = Q_3 \times Q_{8}^2$]{
        \label{fig:1}
        \includegraphics[width=6cm]{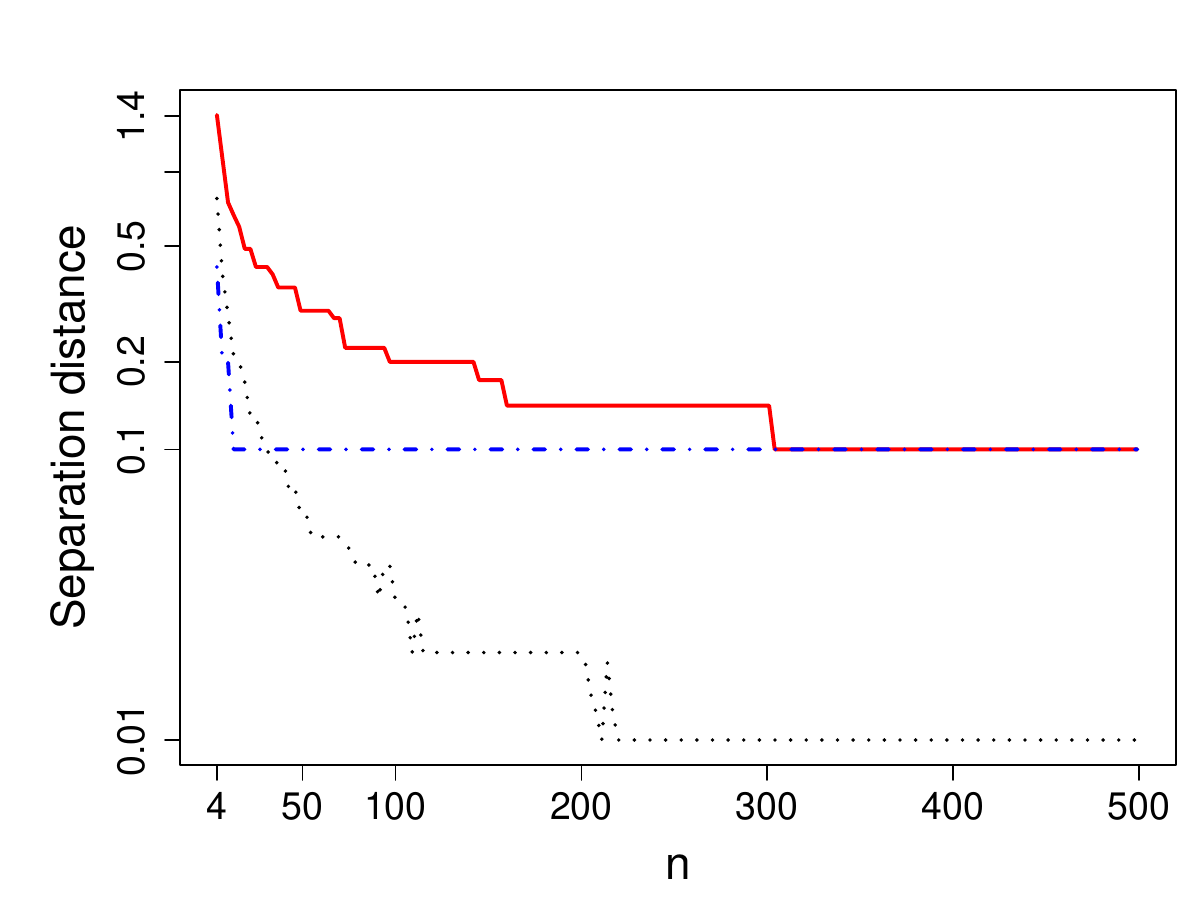}	
      }
      \subfigure[$p=3$, $\calG = Q_8 \times (3/4)Q_3 \times (3/4)^2Q_{8}$ 
          ]{
        \includegraphics[width=6cm]{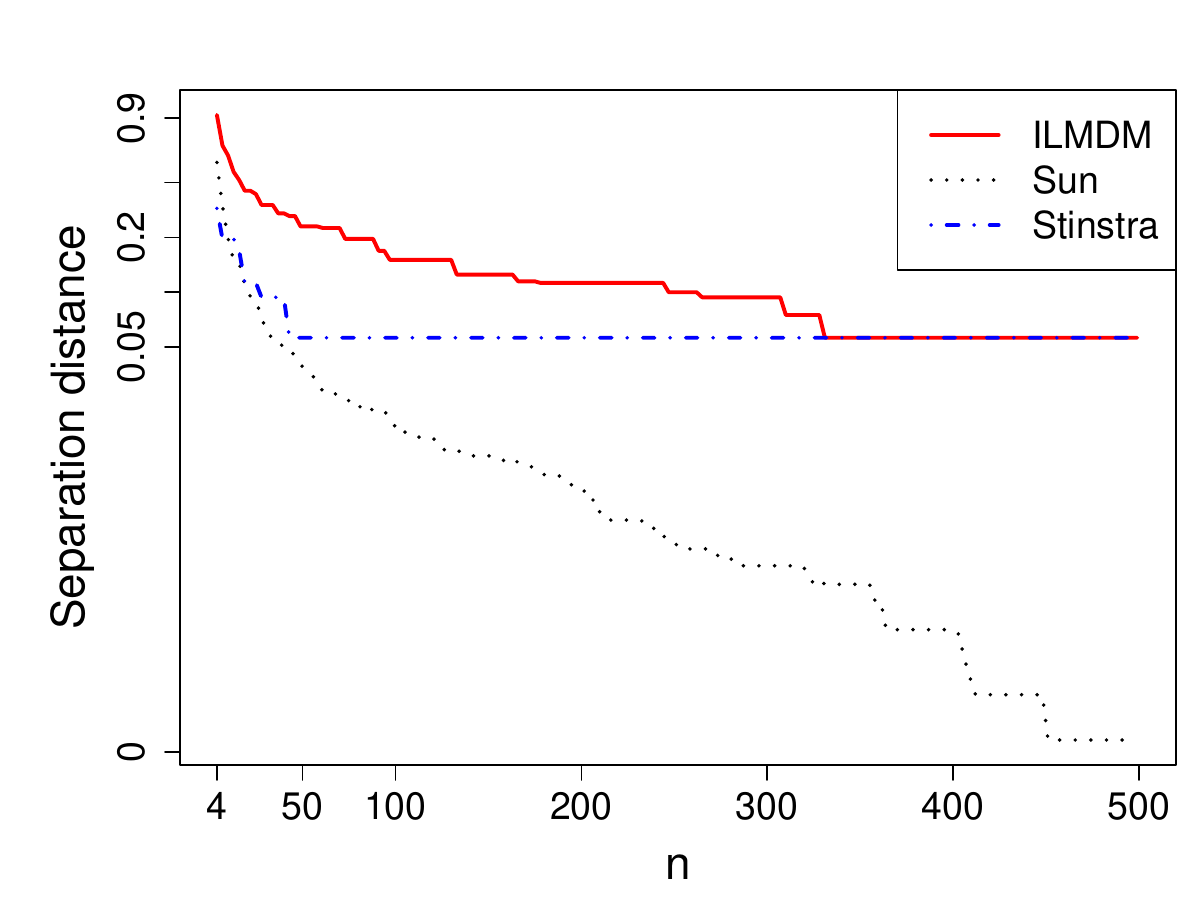}
      }
    \end{minipage}
    \begin{minipage}{1\linewidth}	
      \centering
      \subfigure[$p=6$, $\calG = Q_1 \times Q_3 \times Q_{4} \times Q_6 \times Q_8 \times Q_9$]{
        \label{fig:1}
        \includegraphics[width=6cm]{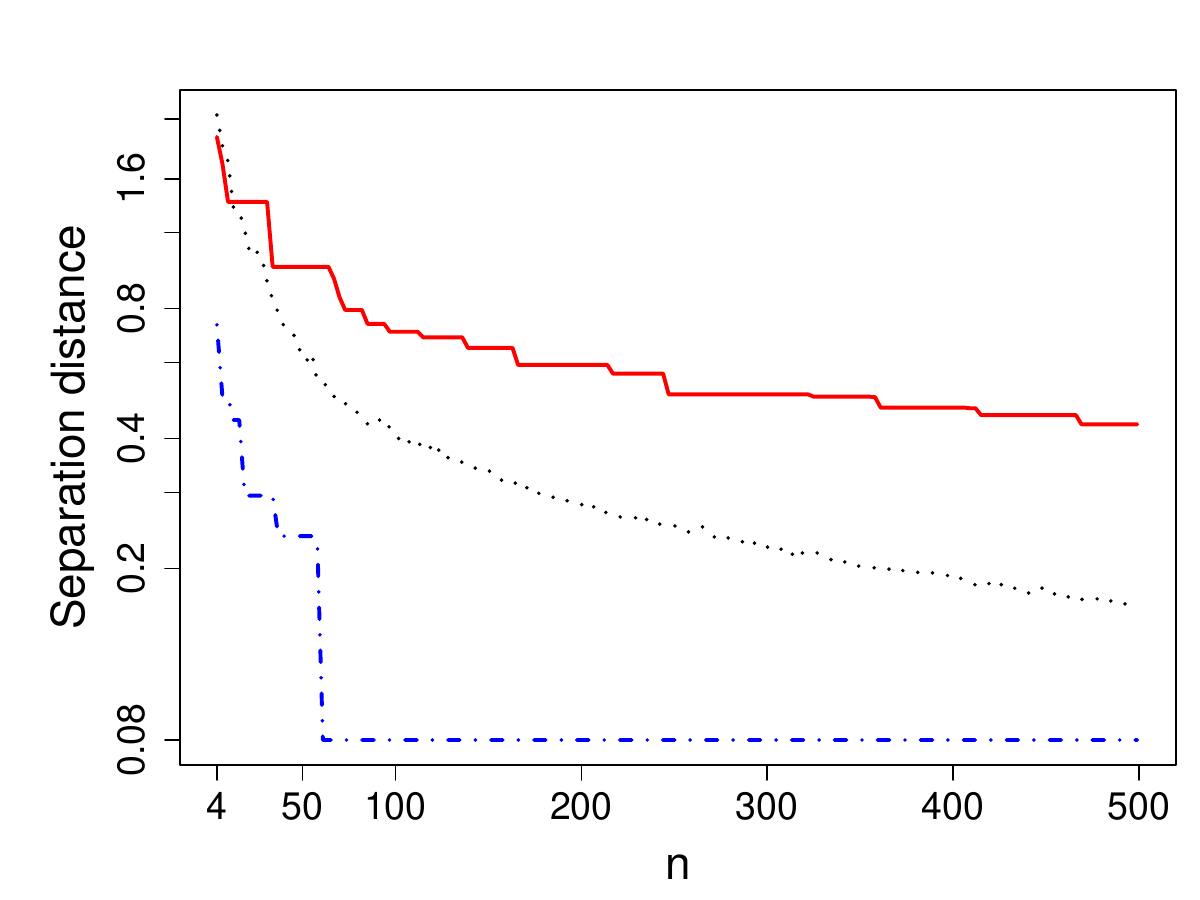}	
      }
      \subfigure[$p{=}15$, $\calG = Q_2 \times (3/4)Q_1 {\times} (3/4)^2Q_{3} \times \prod_{i=3}^{4} (3/4)^{i}Q_1 \times (3/4)^5Q_5 \times \prod_{i=6}^{9} (3/4)^{i}Q_1 \times (3/4)^{10}Q_4 \times (3/4)^{11} Q_2 \times (3/4)^{12}Q_6 \times \prod_{i=13}^{14} (3/4)^{i}Q_1$]{
        \label{fig:2}
        \includegraphics[width=6cm]{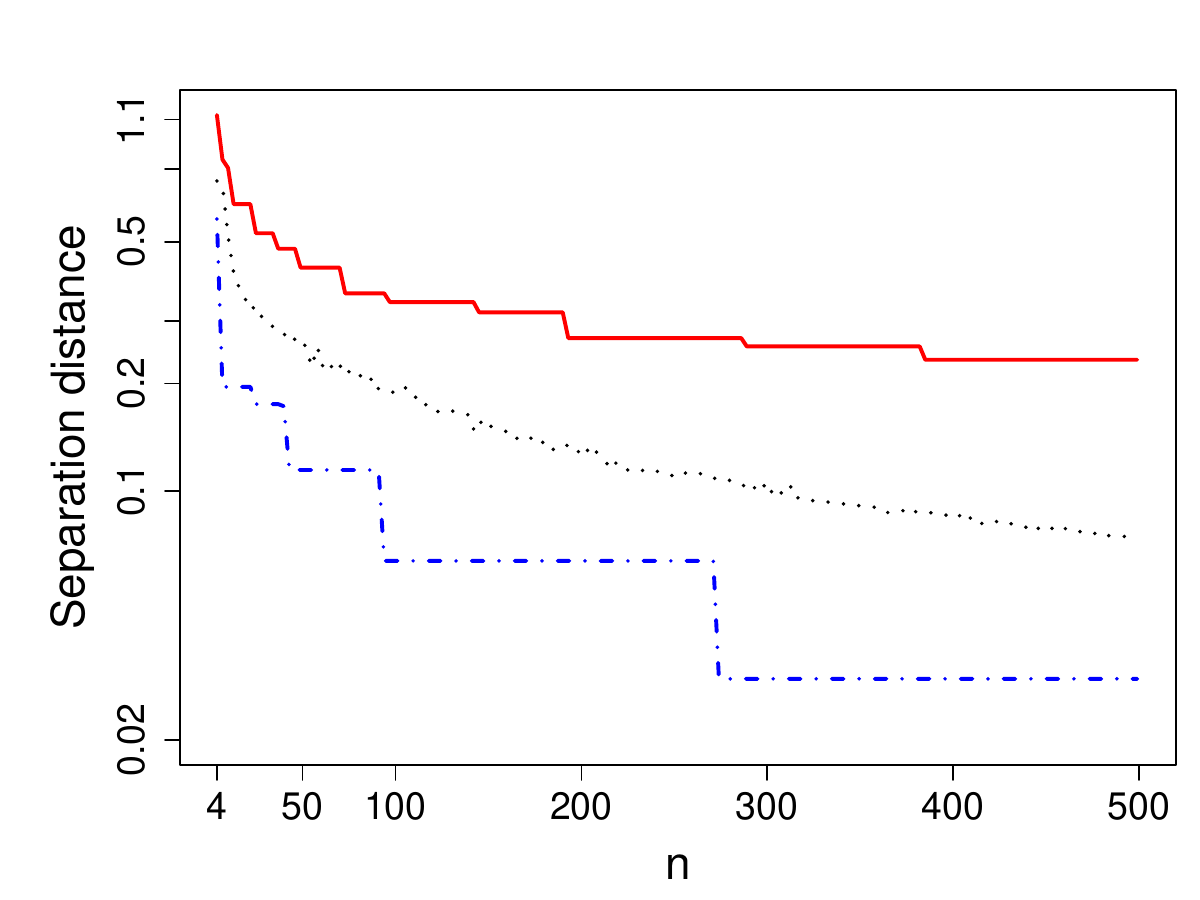}
      }
    \end{minipage}
    \vspace{-1em}
    \caption{Separation distance of three types of maximin distance designs when all variables are ordinal, newly proposed ILMDM (solid), designs by \cite{sun2019synthesizing} (dotted), and designs by \cite{stinstra2003constrained} (dash-dotted).}
    \label{fig:odn}
  \end{figure}

\section{Numerical comparisons}  \label{sec:compare}
In this section, we compare our newly proposed ILMDM with maximin distance designs constructed by \cite{he2019interleaved} (ILMDC), \cite{sun2019synthesizing}, and \cite{stinstra2003constrained}. 
We begin by comparing the methods in scenarios where 
all variables are ordinal. Specifically, we investigate nine types of ordinal variables, as shown in Table~\ref{tab:QQ},  
\CUT{
\begin{align*}
  &Q_1 = \{0,1\}, \\ 
  &Q_2 = \{0,0.5,1\}, \\ 
  &Q_3 = \{0,0.1,0.3,0.6,1\},\\
  &Q_4 = \{0,0.25,0.5,0.75,1\}, \\ 
  &Q_5 = \{0,0.2,0.3,0.5,0.7,1\},\\ 
  &Q_6 = \{0,0.2,0.4,0.6,0.8,1\},\\
  &Q_7 = \{0,0.1,0.3,0.5,0.7,0.9,1\},\\
  &Q_8 = \{0,0.1,0.2,0.3,0.4,0.5,0.6,0.7,0.8,0.9,1\},\\
  &Q_9 = \{0,0.02,0.03,0.05,…,0.83,0.89,0.97,1\},
\end{align*}}
where $Q_9$ consists of 0, 1, and the quotients of all prime numbers less than 100 divided by 100. 
Figure~\ref{fig:odn} presents the results for four scenarios in $3\leq p\leq 15$ with equal or unequal weights.  
Designed for all-continuous circumstances, we do not include ILMDC here. 
Originally, the method of \cite{stinstra2003constrained} generates designs within constrained input spaces inside $[0,1]^p$. 
We extend this method to handle mixed-type variables by translating the allowable level constraints into constraints on the input space.
Given the large differences, the separation distances are plotted on a logarithmic scale, which accentuates the gap between ILMDM and the other methods. 
From the results, ILMDM significantly outperform those from \cite{sun2019synthesizing} and \cite{stinstra2003constrained}, with the exception that those from \cite{stinstra2003constrained} are sometimes equally good as ILMDM. 

  \begin{figure}[t!]
    \centering
    \begin{minipage}{1\linewidth}	
      \centering
      \subfigure[$p=2$, {$\calG = [0,1] \times Q_8$}]{
        \includegraphics[width=6cm]{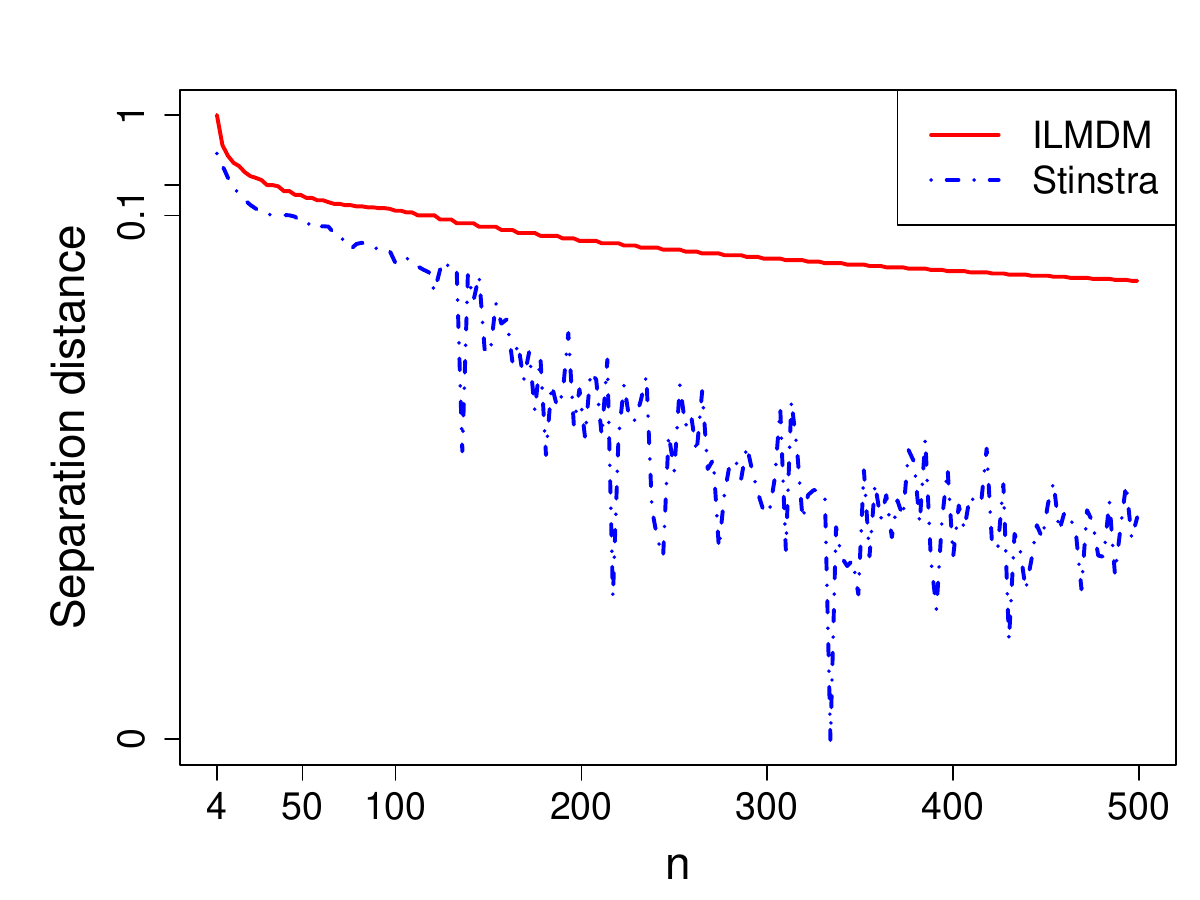}	
      }
      \subfigure[$p=5$, {$\calG =[0,1]^2 \times Q_1 \times Q_5 \times Q_9 $}]{
        \includegraphics[width=6cm]{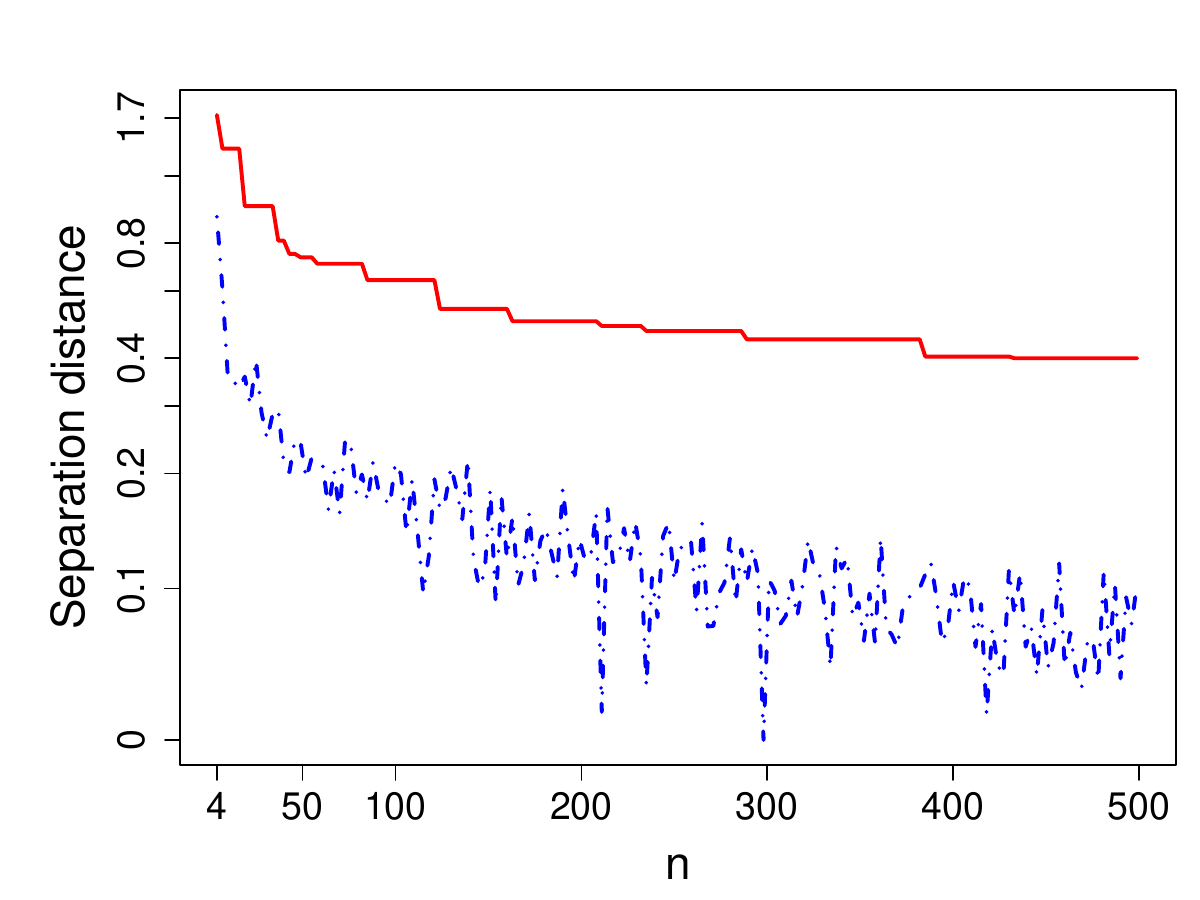}	
      }	
    \end{minipage}
    \begin{minipage}{1\linewidth}	
      \centering
      \subfigure[$p=5$, {$\calG =[0,1] \times (3/4)Q_6 \times [0,(3/4)^2] \times [0,(3/4)^3] \times (3/4)^4Q_7$}]{
        \includegraphics[width=6cm]{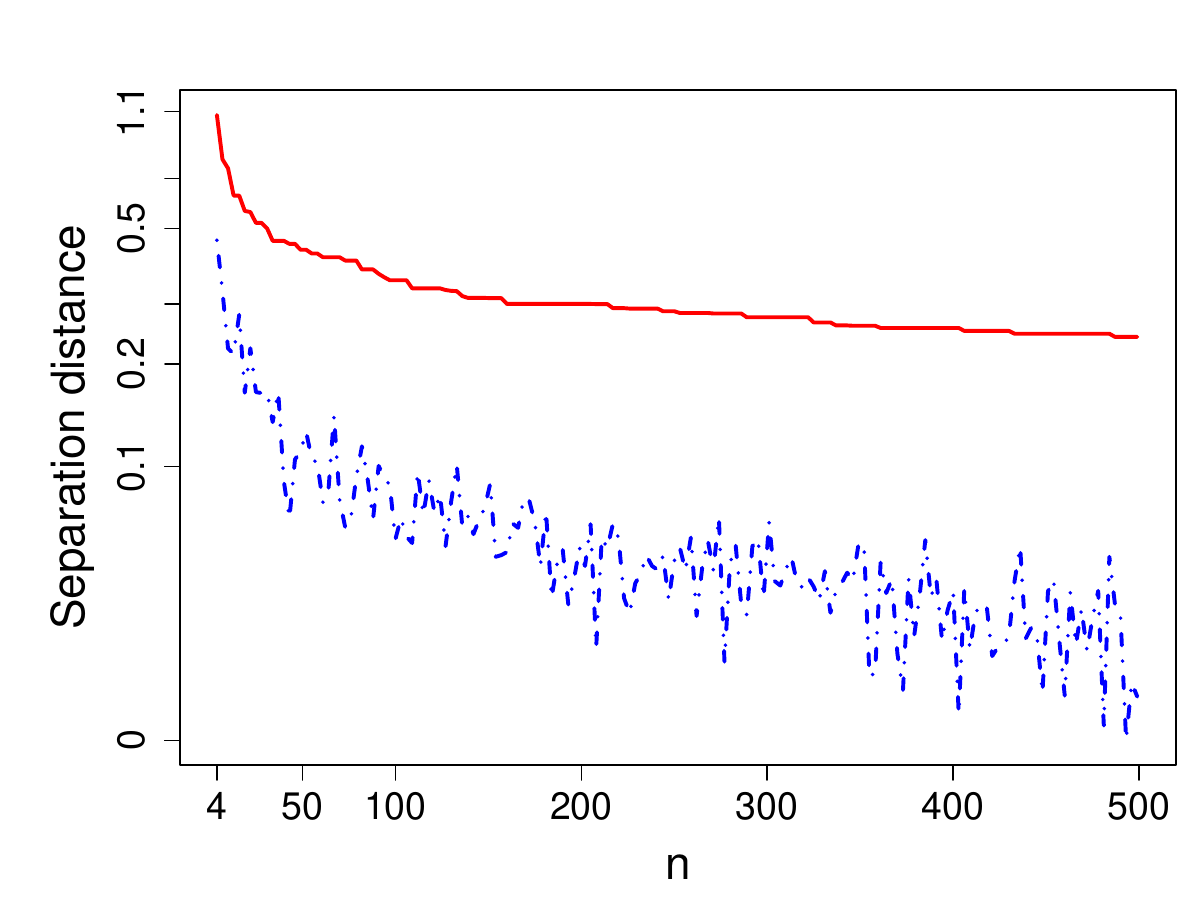}
      }
      \subfigure[$p=8$, {$\calG = [0,1]^4 \times Q_1^4$} 
          ]{
        \includegraphics[width=6cm]{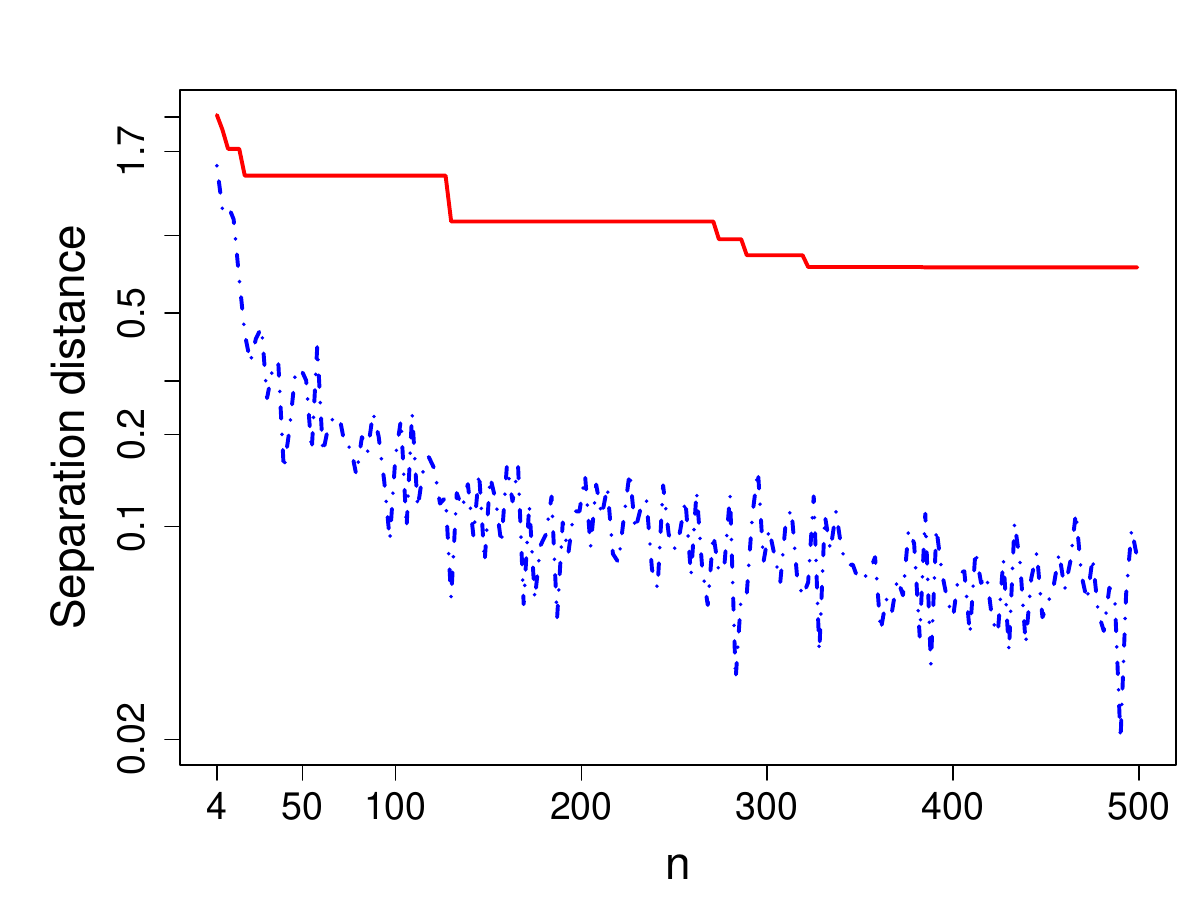}	
      }	
    \end{minipage}
    \begin{minipage}{1\linewidth}	
      \centering
      \subfigure[$p=8$, {$\calG =[0,1]^4 \times Q_1 \times Q_3 \times Q_4 \times Q_8$}]{
        \includegraphics[width=6cm]{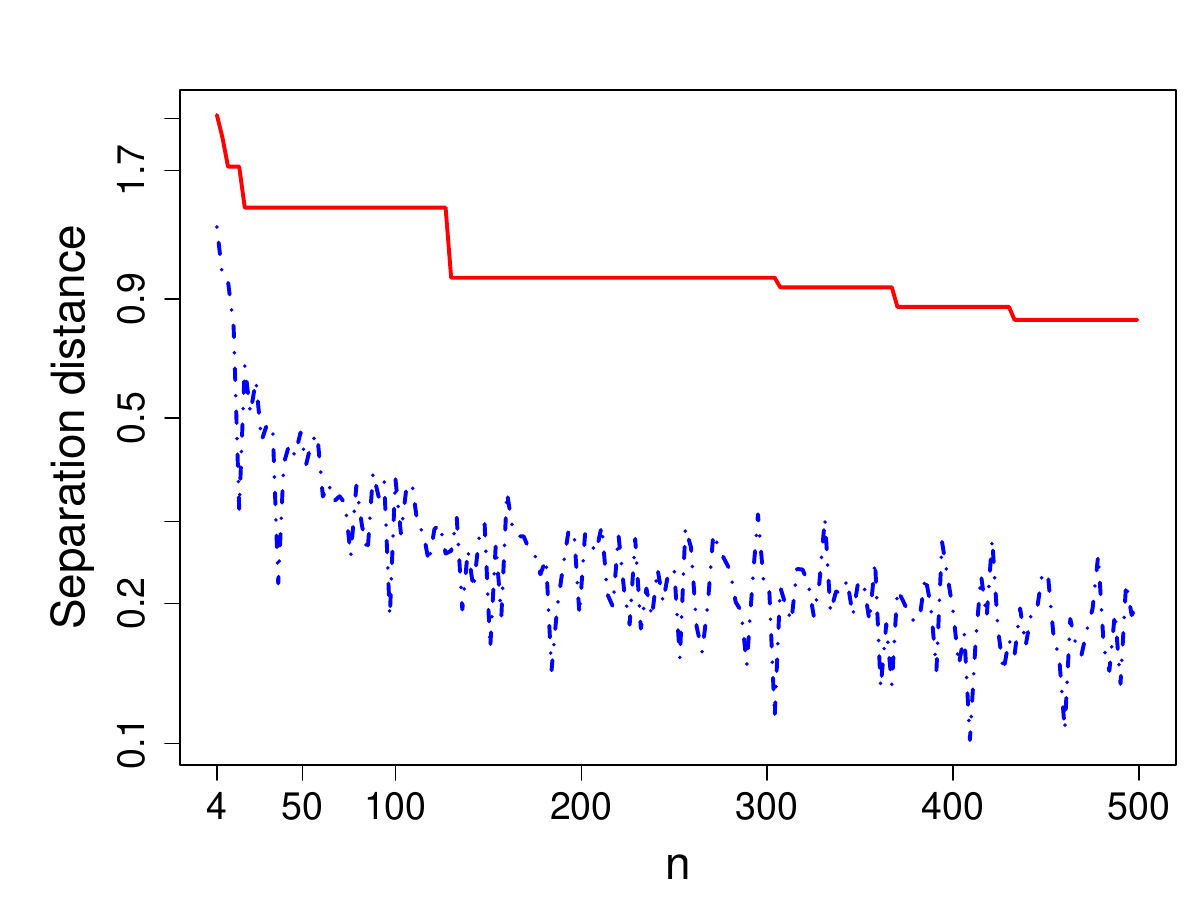}
      }
      \subfigure[$p=12$, {$\calG = [0,1]^3 \times Q_1^3 \times \prod_{i=2}^7 Q_i 
      $}]{
        \includegraphics[width=6cm]{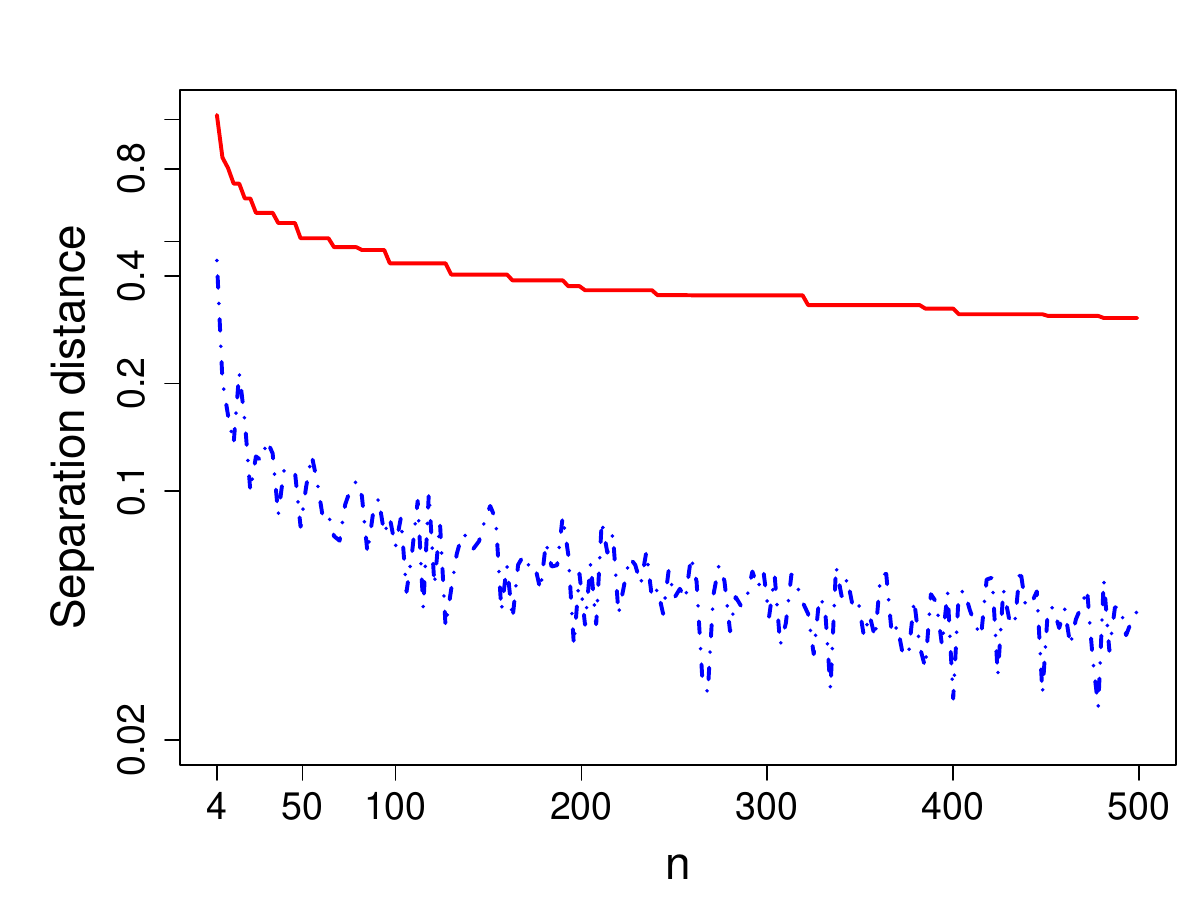}
      }
    \end{minipage}
    \caption{Separation distance of two types of maximin distance designs with mixed continous and ordinal variables, newly proposed ILMDM (solid) and designs by \cite{stinstra2003constrained} (dash-dotted).
    }
    \vspace{-0.2in}		
    \label{fig:con_odn}
  \end{figure}

Next, we examine scenarios involving mixed continuous and ordinal variables. 
Since the method of \cite{sun2019synthesizing} applies to either all-continuous or all-ordinal scenarios, 
The method of \cite{stinstra2003constrained} is the only applicable competitor to ILMDM. 
Figure~\ref{fig:con_odn} presents the results for six scenarios in $2\leq p\leq 12$ with equal or unequal weights. 
From the results, ILMDM exhibits consistently superior separation distance that are typically twice to ten times as high as designs of \cite{stinstra2003constrained}.

\begin{figure}[t!]
  \centering
  \begin{minipage}{1\linewidth}	
    \centering
    \subfigure[$p=3, {\calG=[0,1]^3}$]{
      \label{fig:1}
      \includegraphics[width=6cm]{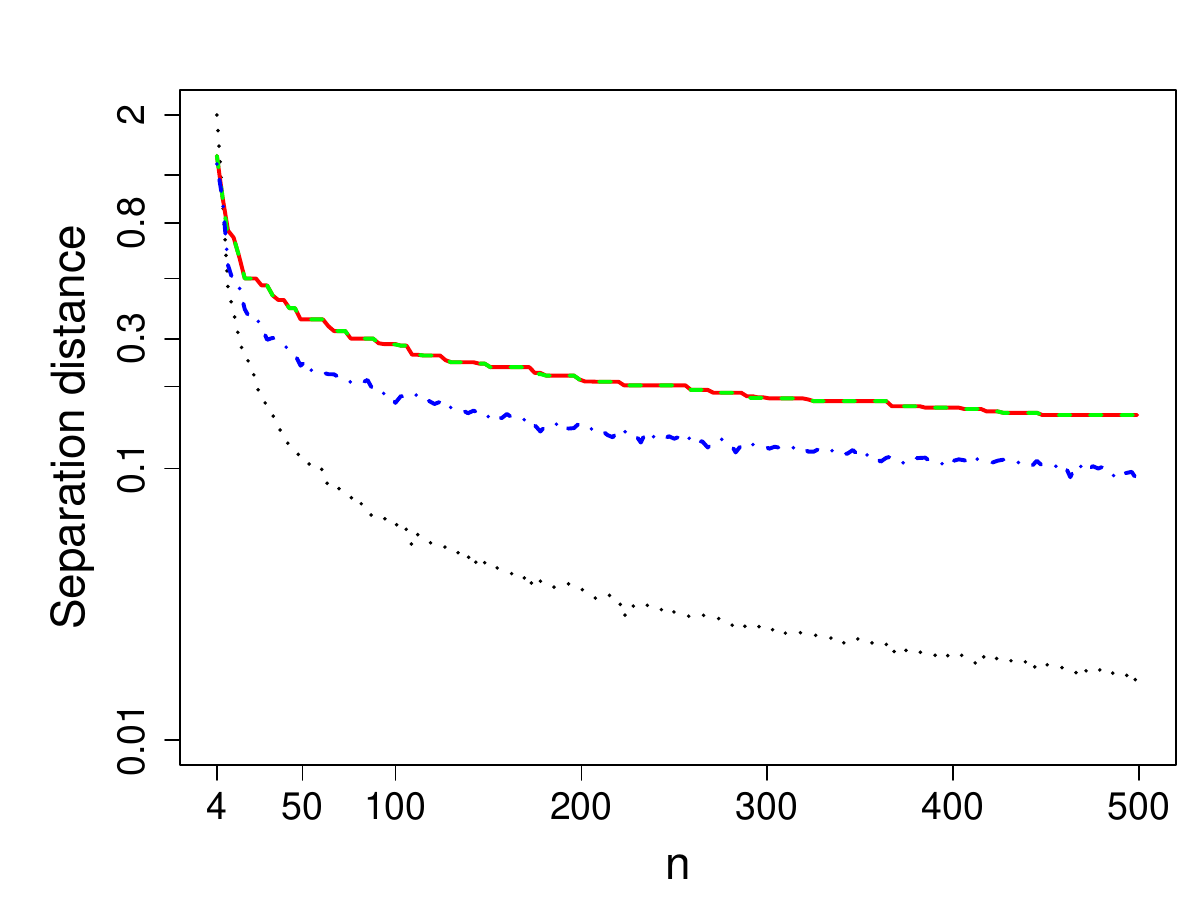}	
    }
    \subfigure[$p=15, {\calG= \prod_{k=1}^{15}[0,(3/4)^{k-1}]}$]{
      \label{fig:2}
      \includegraphics[width=6cm]{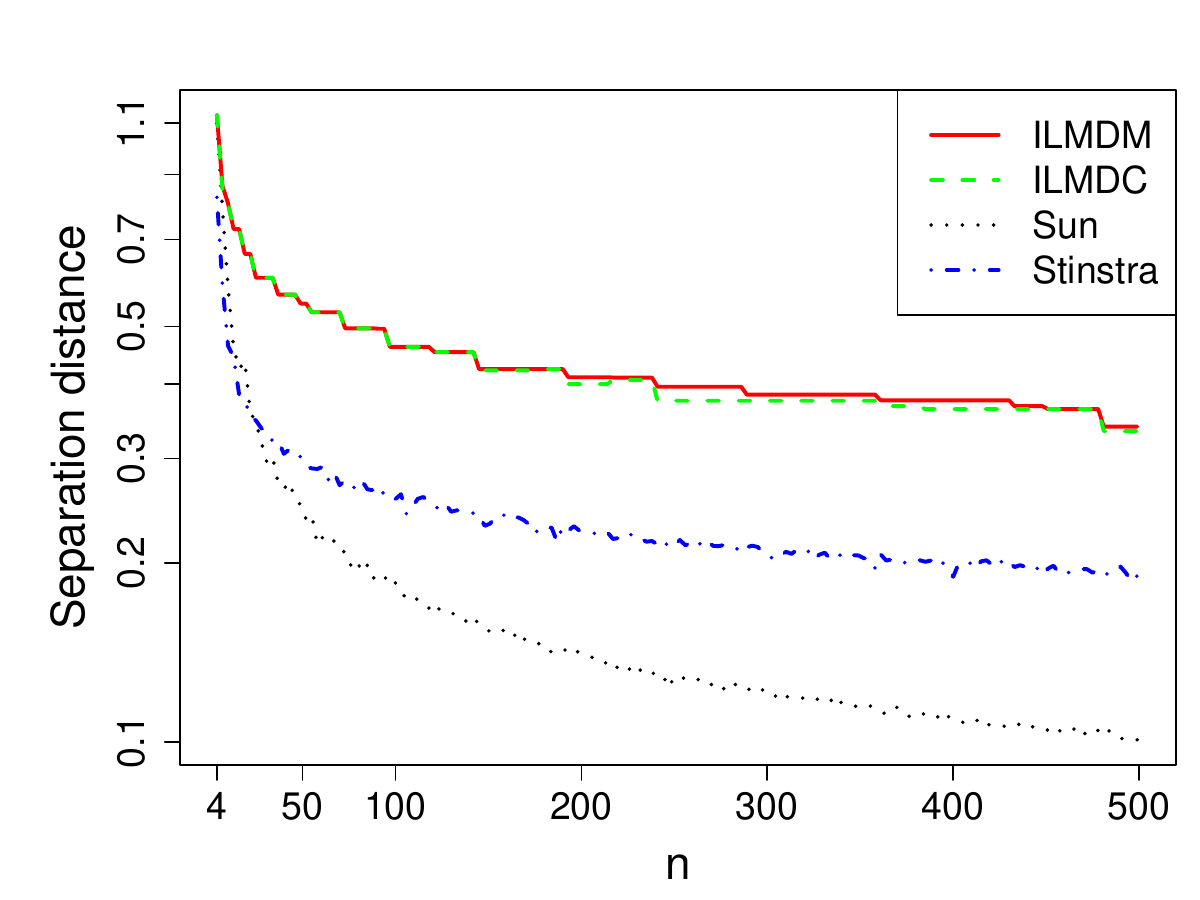}
    }
  \end{minipage}
  \caption{Separation distance of four types of maximin distance designs when all variables are continuous, newly proposed ILMDM (solid), ILMDC by \citet{he2019interleaved} (dashed), designs by \citet{sun2019synthesizing} (dotted), and designs by \cite{stinstra2003constrained} (dash-dotted). 
  }
  \label{fig:con}
\end{figure}

Finally, we consider the cases with all continuous variables. 
Figure~\ref{fig:con} presents the results for $p=3$ with equal weights and $p=15$ with unequal weights. 
From the results, while ILMDM and ILMDC yield nearly identical separation distance in the majority of cases, there are occasional instances where ILMDM slightly outperforms ILMDC. 
The methods by \cite{sun2019synthesizing} and \cite{stinstra2003constrained} are substantially not as effective as ILMDC and ILMDM. 

To conclude, our proposed designs demonstrate excellent performance across all cases involving continuous, ordinal, and binary variables.




\appendix

\section*{S1: Proofs of theorems}\label{appn} 

\subsection*{Preliminary Notation and Omitted Proofs for Theorems~\ref{thm:0_p:m}-\ref{thm:m:r}}
For a point $\bm{x} = (x_1, \ldots, x_p) \in \mathbb{R}^p$ and a set ${\cal A} \subset \{1,\ldots,p\}$. 
Let 
\[\bm{x}_{\cal A} = \left\{(x_{k_1}, \ldots, x_{k_h}) : k_i \text{ is the $i$-th smallest number in } {\cal A}, i = 1, \ldots, m(\calA) \right\},\] 
\ie $\bm{x}_{\cal A}$ consists of the dimensions in ${\cal A}$ of the point $\bm{x}$. 
Theorems~\ref{thm:0_p:m}-\ref{thm:m:r}  can be proved in the same way as their conterparts in \citet{he2019interleaved} and thus we omit the proofs. 

\subsection*{Proof of Theorem~\ref{thm:0_p:u}}
\begin{proof}
Clearly, there exist $\bm{a},\bm{b} \in ( \calL \oplus \bm{u} ) \cap \prod_{k=1}^p [0,s_k-1]$, $\bm{a} \neq \bm{b}$, 
and 
\[ \rho\left\{\calD(\calL,\bm{s},\caly,\bm{u})\right\} = d\left(\caly_{(\bm{a})},\caly_{(\bm{b})}\right)  = \left\{ \sum_{k=1}^p \left( \caly_{k(b_k)} - \caly_{k(a_k)} \right)^2 \right\} .\] 

Suppose that there is a $k$ such that $|b_k-a_k|>2$. 
Without lose of generality assume $b_k>a_k$. 
Then $\bm{b}-2\bm{e}_k \in ( \calL \oplus \bm{u} ) \cap \prod_{k=1}^p [0,s_k-1]$, $\bm{a} \neq \bm{b}-2\bm{e}_k$, 
and $d\left(\caly_{(\bm{a})},\caly_{(\bm{b}-2\bm{e}_k)}\right)  < \rho\{\calD(\calL,\bm{s},\caly,\bm{u})\}$,  
which contradicts to \eqref{eqn:ILDM}. 
Consequently, the assumption that there is a $k$ such that $|b_k-a_k|>2$ cannot hold. 

Suppose that there is a $k$ such that $|b_k-a_k|=2$ and there is a $j\neq k$ such that $b_j\neq a_j$. 
Without lose of generality assume $b_k>a_k$. 
Then $\bm{b}-2\bm{e}_k \in ( \calL \oplus \bm{u} ) \cap \prod_{k=1}^p [0,s_k-1]$, $\bm{a} \neq \bm{b}-2\bm{e}_k$, 
and $d\left(\caly_{(\bm{a})},\caly_{(\bm{b}-2\bm{e}_k)}\right)  < \rho\{\calD(\calL,\bm{s},\caly,\bm{u})\}$,  
which contradicts to \eqref{eqn:ILDM}. 
Consequently, the assumption that there is a $k$ such that $|b_k-a_k|=2$ and there is a $j\neq k$ such that $b_j\neq a_j$ cannot hold. 

Consider three cases on $\bm{a}$ and $\bm{b}$. 
Firstly, when $\bm{b} = \bm{a} + 2\bm{e}_k$ for a $k$. 
Then $s_k \geq 3$ and $d\left(\caly_{(\bm{a})},\caly_{(\bm{b})}\right) \geq d^+(\caly_k)$. 
On the other hand, there exists a $0\leq z\leq s-3$ such that $\caly_{k(z+2)}-\caly_{k(z)}=d^+(\caly_k)$. 
Then $z\bm{e}_k,(z+2)\bm{e}_k \in ( \calL \oplus \bm{0}_p ) \cap \prod_{k=1}^p [0,s_k-1]$ 
and  \[ d\left(\caly_{(z\bm{e}_k)},\caly_{((z+2)\bm{e}_k)} \right) = d^+(\caly_k) .\] 
Therefore, $\rho\{\calD(\calL,\bm{s},\caly,\bm{0}_p)\} \leq \rho\{\calD(\calL,\bm{s},\caly,\bm{u})\}$. 

Secondly, when $\bm{b} = \bm{a} - 2\bm{e}_k$ for a $k$. 
Similarly, we have $\rho\{\calD(\calL,\bm{s},\caly,\bm{0}_p)\} \leq \rho\{\calD(\calL,\bm{s}, \allowbreak \caly, \allowbreak \bm{u})\}$. 

Thirdly, when $|b_k-a_k|\leq 1$ for any $k$. 
For any $k$ such that $u_k$ is even, let $\tilde a_k = a_k$ and $\tilde b_k = b_k$. 
For any $k$ such that $u_k$ is odd, let $\tilde a_k = b_k$ and $\tilde b_k = a_k$. 
Then $\bm{\tilde a} = (\tilde a_1,\ldots,\tilde a_p) \in \prod_{k=1}^p [0,s_k-1]$ and $\bm{\tilde b} = (\tilde b_1,\ldots,\tilde b_p) \in \prod_{k=1}^p [0,s_k-1]$.  
For any $k$, because $|b_k-a_k|\leq 1$, both $\tilde a_k-a_k$ and $\tilde b_k - b_k$ are even integers. 
Because $\calL$ is a standard interleaved lattice, $2\bm{e}_k \in \calL$ for any $k$. 
Therefore, $\bm{\tilde a} \in \calL$ and $\bm{\tilde b} \in \calL$.  
Consequently, $\bm{\tilde a} \in ( \calL \oplus \bm{0}_p ) \cap \prod_{k=1}^p [0,s_k-1]$ and $\bm{\tilde b} \in ( \calL \oplus \bm{0}_p ) \cap \prod_{k=1}^p [0,s_k-1]$.  
On the other hand, $|\caly_{k(a_k)}-\caly_{k(b_k)}| = |\caly_{k(\tilde a_k)}-\caly_{k(\tilde b_k)}|$ for any $k$. 
Therefore, \[ \rho\{\calD(\calL,\bm{s},\caly,\bm{0}_p)\} \leq d\left(\caly_{(\bm{\tilde a})},\caly_{(\bm{\tilde b})}\right) \leq \rho\{\calD(\calL,\bm{s},\caly,\bm{u})\}. \]

Combining the three cases on $\bm{a}$ and $\bm{b}$, we have 
\[ \rho\{\calD(\calL,\bm{s},\caly,\bm{0}_p)\} \leq \rho\{\calD(\calL,\bm{s},\caly,\bm{u})\}. \]
Using similar arguments, we can show that 
\[ \rho\{\calD(\calL,\bm{s},\caly,\bm{u})\} \leq \rho\{\calD(\calL,\bm{s},\caly,\bm{0}_p)\}, \]
which completes the proof.
\end{proof}

\subsection*{Proof of Theorem~\ref{thm:sep}}
\begin{proof}

Suppose $\bar{\calD} = {\calD} (\bar{\calL}, \bar{\bm{s}}, \bar\caly,\bm{0}_p) $ is a design generated via (\ref{eqn:ILDM}) with $m(\bar{\calD}) \geq n$, $\bar{\calL}$ being a standard interleaved lattice, $\bar \caly \subset \calG$, and $\bar{\bm{s}} = (\bar{s}_1, \ldots,\bar{s}_p)$ with $\bar{s}_k =1$ for at least a $k \in \{1,\ldots,p\}$. 
We shall show that there exist a design $\tilde{\calD}$ in (\ref{eqn:ILDM}) with $m(\tilde{\calD}) \geq n$, $\tilde{\calL}$ being a standard interleaved lattice, $\bm{\tilde u}=\bm{0}_p$, $\bm{\tilde s} = (\tilde s_1, \ldots, \tilde s_p)$ with $\tilde s_k \ge 2$ for all $k$, and  $\rho(\tilde{\calD}) \ge \rho(\bar{\calD})$.

Let ${\cal K}_1 = \{k : \bar{s}_k =1\}$, ${\cal K}_2 =\{k: \bar{s}_k \ge2 \}$, $h=m({\cal K}_2)$, and 
$\bar{\calL}_{{\cal K}_2} = \{x_{{\cal K}_2} \in \mathbb{Z}^h : x \in \bar{\calL}\}$. 
Considering two cases on $\bar{\calL}_{{\cal K}_2}$. 
Firstly, if $\bar{\calL}_{{\cal K}_2} = \mathbb{E}^h$. 
Let $\hat{\calL} = \mathbb{Z}^p$. 
For $k \in {\cal K}_2$, let $\hat{s}_k = \lceil \bar{s}_k/2 \rceil$ and 
$\hat{y}_{k,i} = \bar{y}_{k,2i-1}$ for $i =1 , \ldots, \hat{s}_k$. 
For $k \in {\cal K}_1$, let $\hat{s}_k = \bar{s}_k$ and $\hat{y}_{k,i} = \bar{y}_{k,i}$ for $i =1 , \ldots, \hat{s}_k$. 
Secondly, if $\bar{\calL}_{{\cal K}_2} \neq \mathbb{E}^h$, let 
 \begin{align*}
 \hat{\calL} = \left\{ \bm{x} = (x_1, \ldots, x_p): \bm{x}_{\calK_2} \in \bar{\calL}_{{\cal K}_2}, \bm{x}_{{\cal K}_1} \in  \mathbb{Z}^{p-h}  \right\} ,
 \end{align*}
$\hat{s}_k = \bar{s}_k$, and $\hat{y}_{k,i} = \bar{y}_{k,i}$ for $i =1 , \ldots, \hat{s}_k$ and $k = 1, \ldots, p $. 
For both cases on $\bar{\calL}_{{\cal K}_2}$, let $\hat {\caly}_k = \{\hat{y}_{k,i} : i=1,\ldots,\hat s_i\}$ for any $k$, $\hat{\caly} = \prod_{k=1}^p \hat{\caly}_k$, and $\hat{\calD} = {\calD} (\hat{\calL}, \hat{\bm{s}}, \hat\caly,\bm{0}_p)$. 
Then the $\hat{\calL}$ is a standard interleaved lattice, $\rho(\hat{\calD}) = \rho(\bar{\calD})$, and $m(\hat{\calD}) = m(\bar{\calD})$. 

Next, let $\hat{{\cal K}}_1 = \{k : \hat{s}_k =1\}$ and $\hat{{\cal K}}_2 =\{k: \hat{s}_k \ge2 \}$ with ${\hat{h}} = m(\hat{{\cal K}}_2)$, and
\[\hat{\calL}_{\hat{{\cal K}}_2} = \left\{\bm{x}_{\hat{{\cal K}}_2} \in \mathbb{Z}^{\hat{h}} : \bm{x}   \in \hat{\calL} \right\}. \]
Obviously, $\hat{\calL}_{\hat{{\cal K}}_2}$ is a lattice, $\mathbb{E}^{\hat{h}} \subset \hat{\calL}_{\hat{{\cal K}}_2} \subset \mathbb{Z}^{\hat{h}}$, and $\hat{\calL}_{\hat{{\cal K}}_2} \neq \mathbb{E}^{\hat{h}}$. Partition $\hat{\calL}_{\hat{{\cal K}}_2}$ into two subsets, $\hat{\calL}_{\hat{{\cal K}}_2}^{(1)}$ and
$\hat{\calL}_{\hat{{\cal K}}_2}^{(2)}$, such that $\mathbb{E}^{\hat{h}} \subset \hat{\calL}_{\hat{{\cal K}}_2}^{(1)} \subset \mathbb{Z}^{\hat{h}}$ and $\hat{\calL}_{\hat{{\cal K}}_2}^{(2)}$ is a translation of $\hat{\calL}_{\hat{{\cal K}}_2}^{(1)}$. To do so, assume $\hat{\calL}_{\hat{{\cal K}}_2} \cap \{0,1\}^{\hat{h}} $ has $2^q$ elements.
Because $\hat{\calL}_{\hat{{\cal K}}_2} \neq \mathbb{E}^{\hat{h}} $, we have $q \ge 1$. Select arbitrary $q-1$ elements of $\hat{\calL}_{\hat{{\cal K}}_2} \cap \{0,1\}^{\hat{h}} $, $v_1, \ldots, v_{q-1}$, such that $\sum_{i=1}^{q-1} c_i v_i \notin \mathbb{E}^{{\hat{h}}}$
 for any $(c_1, \ldots, c_{q-1}) \in \{0,1\}^{q-1} \setminus \{0_{{\hat{h}}} \} $. Letting $\hat{\calL}_{\hat{{\cal K}}_2}^{(1)}$ be the space generated from $v_1, \ldots, v_{q-1}, 2e_1, \ldots, 2e_{{\hat{h}}}$, and $\hat{\calL}_{\hat{{\cal K}}_2}^{(2)} = \hat{\calL}_{{\cal K}_2} \setminus \hat{\calL}_{\hat{{\cal K}}_2}^{(1)}$, the $\hat{\calL}_{\hat{{\cal K}}_2}^{(1)}$ and $\hat{\calL}_{\hat{{\cal K}}_2}^{(2)}$ satisfy our requirements. Let
 \begin{align*}
  \tilde{\calL}_1 &= \left\{ \bm{x} = (x_1, \ldots, x_p): \bm{x}_{\hat{K}_2} \in \hat{\calL}^{(1)}_{\hat{{\cal K}}_2}, \bm{x}_{{\cal K}_1} \in \mathbb{E}^{p-\hat{h}}  \right\} ,\\
 \tilde{\calL}_2 &= \left\{ \bm{x} = (x_1, \ldots, x_p): \bm{x}_{\hat{{\cal K}}_2} \in \hat{\calL}^{(2)}_{\hat{{\cal K}}_2}, \bm{x}_{{\cal K}_1} \in (\mathbb{E} \oplus \bm{1})^{p-\hat{h}}  \right\} ,
 \end{align*}
 and $\tilde{\calL} = \tilde{\calL}_1 \cup \tilde{\calL}_2$. Then $\tilde{\calL}$ is a standard interleaved lattice. 
Let $\tilde{s}_k = \hat{s}_k$ and $\tilde{\caly}_k = \bar{\caly}_k$ for $k \in \hat{{\cal K}}_2$. 
Let $\tilde{s}_k = 2$, $\tilde{\caly}_{k} = \{0,w_k\}$ for $k \in \hat{{\cal K}}_1$. 
Furthermore, let $\tilde {\caly} = \prod_{k=1}^p \tilde{\caly}_k$ and $\tilde{\calD} = {\calD} (\tilde{\calL}, \tilde{\bm{s}},  \tilde\caly,\bm{0}_p)$. 
Then $\rho(\tilde{\calD}) \ge \rho(\hat{\calD})$ and $m(\tilde{\calD}) = m(\hat{\calD})$. 

In sum, for an arbitrary design $\bar{\calD} = {\calD} (\bar{\calL}, \bar{\bm{s}}, \bar\caly, \bm{0}_p)$ with $\bar{s}_k =1$ for some $k$s, there will be a corresponding design $\tilde{\calD} = {\calD} (\tilde{\calL}, \tilde{\bm{s}},  \tilde\caly, \bm{0}_p)$
such that  $\tilde{s}_k \ge 2$ for all $k$, $\rho(\tilde{\calD}) \ge \rho(\bar{\calD})$, and $m(\tilde{\calD}) = m(\bar{\calD}) \ge n$.
This verifies that for any fixed $\calG \subset \prod_{k=1}^p [0, w_k]$ and $n\geq 2$, among all interleaved
lattice-based maximin distance designs for mixed-type variables $\calD$ in \eqref{eqn:ILDM} with $m(\calD)\geq n$, the highest separation distance $\rho(\calD)$ can be attained by a design with $\bm{u}=\bm{0}_p$ and $s_k \ge 2$ for all $1\leq k \leq p$.
Clearly, from switching the lowest element of $\tilde{\caly}_k$ to be zero and the highest element of $\tilde{\caly}_k$ to be $w_k$ for any $k$, the size of the design remains unchanged and the separation distance does not decrease. 
Therefore, for any fixed $\calG \subset \prod_{k=1}^p [0, w_k]$ and $n\geq 2$, among all interleaved lattice-based maximin distance designs for mixed-type variables $\calD$ in \eqref{eqn:ILDM} with $m(\calD)\geq n$, the highest separation distance $\rho(\calD)$ can be attained by a design with $\bm{u}=\bm{0}_p$ and $s_k \ge 2$, $\caly_{k(1)}=0$, and $\caly_{k(s_k)}=w_k$ for all $1\leq k \leq p$.

    Suppose $\calD = \calD(\calL,\bm{s},\caly,\bm{0}_p)$ in (\ref{eqn:ILDM}) is one such optimal design. 
Let 
\[ \calI = \left( \calL \cap \prod_{k=1}^p [0,s_k-1] \right) \oplus \bm{1}_p. \] 
Then 
    \begin{align*}
    \rho(\calD) = \min_{\bm{a},\bm{b} \in \calI } d\left(\caly_{(\bm{a})},\caly_{(\bm{b})}\right) 
 = \min_{\bm{a},\bm{b} \in \calI } \left\{ \sum_{k=1}^p \left(\caly_{k(a_k)} - \caly_{k(b_k)}\right)^2 \right\}^{1/2} .
    \end{align*}
    
    
Firstly, suppose $\bm{\tilde x} = (\tilde x_1,\ldots,\tilde x_p) \in \calH$, $\bm{\tilde x} \neq \bm{0}_p$, and 
\[ \min_{\bm{x} \in \calH, \bm{x} \neq \bm{0}_p} \left\{ \sum_{x_k=1} d^*(\caly_k)^2 \right\}^{1/2} = \left\{ \sum_{\tilde x_k=1} d^*(\caly_k)^2 \right\}^{1/2}. \]
For any $k$, there exists an $1 \leq a_k \leq s_k-1$ such that $\caly_{k(a_k+1)} - \caly_{k(a_k)} = d^*(\caly_k)$. 
Let $\bm{b} = \bm{a} + \bm{\tilde x}$. 
Then $\bm{a},\bm{b} \in \calI$ and \[  \left\{ \sum_{k=1}^p \left(\caly_{k(a_k)} - \caly_{k(b_k)}\right)^2 \right\}^{1/2} = \left\{ \sum_{\tilde x_k=1} d^*(\caly_k)^2 \right\}^{1/2}. \]
Consequently, 
\[ \rho(\calD) \le \min_{\bm{x} \in \calH, \bm{x} \neq \bm{0}_p} \left\{ \sum_{x_k=1} d^*(\caly_k)^2 \right\}^{1/2}.\]

Secondly, suppose $s_k >2$ and $d^+(\caly_k) = \min_{s_j \ge 2}  d^+(\caly_j)$. 
Then there exists an $1 \leq a_k \leq s_k-2$ such that $\caly_{k(a_k+2)} - \caly_{k(a_k)} = d^+(\caly_k)$. 
Let $\bm{b} = \bm{a} + 2\bm{e}_k$. 
Then $\bm{a},\bm{b} \in \calI$ and \[ d\left(\caly_{(\bm{a})},\caly_{(\bm{b})}\right) =  d^+(\caly_k).\]
Consequently, \[\rho(\calD) \le \min_{s_k \ge 2}  d^+(\caly_k).\]
Therefore,
    \begin{equation*}
    \rho(\calD) \le \min\left[ \min_{\bm{x} \in L \cap \{0,1\}^p, \bm{x} \neq \bm{0}_p} \left\{ \sum_{x_k=1} d^*(\caly_k)^2 \right\}^{1/2}, \min_{s_k>2}  d^+(\caly_k)  \right].  
    \end{equation*} 
    
    Conversely, suppose $\rho(\calD) = d\big(\caly_{(\bm{a})},\caly_{(\bm{b})}\big)$, $ \bm{a}, \bm{b} \in \calI$, and  $\bm{a} \neq \bm{b}$. 
Consider two cases on the relationship between $\bm{a}$ and $\bm{b}$. 
Firstly, when $|a_k - b_k| \le 1$ for all $k$. 
Then 
    \[d\left(\caly_{(\bm{a})},\caly_{(\bm{b})}\right)  =  \left\{\sum_{k=1}^p \left(\caly_{k(b_k)} - \caly_{k(a_k)} \right)^{2} \right\}^{1/2} 
 \ge \left\{ \sum_{b_k \neq a_k} d^*(\caly_k)^2 \right\}^{1/2}. \]
Let $\bm{\bar x} = \bm{b} - \bm{a}$. Because $\calL$ is a lattice, $\bm{\bar x} \in \calL$. 
Let $\bm{\tilde x} = (|\bar x_1|, \ldots, |\bar x_p|)$. 
For any $k$, because $|a_k - b_k| \le 1$, $\tilde x_k$ is either $\bar x_k$ or $\bar x_k+2$. 
Because $\calL$ is a standard interleaved lattice, $2\bm{e}_k \in \calL$ for any $k$. 
Therefore, $\bm{\tilde x} \in \calL$. Also because $\bm{\tilde x} \in \{0,1\}^p$, $\bm{\tilde x} \in \calH $. 
On the other hand, \[ \left\{ \sum_{b_k \neq a_k} d^*(\caly_k)^2 \right\}^{1/2} = \left\{ \sum_{\tilde x_k=1} d^*(\caly_k)^2 \right\}^{1/2} \ge  \min_{\bm{x} \in \calH, \bm{x} \neq \bm{0}_p} \left\{ \sum_{x_k=1} d^*(\caly_k)^2 \right\}^{1/2}. \]
Consequently, 
    \[ d\left(\caly_{(\bm{a})},\caly_{(\bm{b})}\right) \ge  \min_{\bm{x} \in \calH, \bm{x} \neq \bm{0}_p} \left\{ \sum_{x_k=1} d^*(\caly_k)^2 \right\}^{1/2}.\] 
    Secondly, assume $|a_k - b_k| \ge 2$ for some $k$. Then there exists a $k$ such that $|a_k - b_k| \ge 2$ and  $s_k >2$. 
Then 
\[  d\left(\caly_{(\bm{a})},\caly_{(\bm{b})}\right) 
 =  \left\{ \sum_{j=1}^p \left(\caly_{j(a_j)} - \caly_{j(b_j)}\right)^2 \right\}^{1/2}  
 \geq \left|\caly_{k(a_k)} - \caly_{k(b_k)}\right| 
 \geq d^+(\caly_k). \]
Consequently, \[d\left(\caly_{(\bm{a})},\caly_{(\bm{b})}\right) \ge  \min_{s_k \ge 2} d^+(\caly_k).\]
Combining the two cases, we have
    \begin{equation*}
    \rho(\calD) \ge \min\left[ \min_{\bm{x} \in \calH, \bm{x} \neq \bm{0}_p} \left\{ \sum_{x_k=1} d^*(\caly_k)^2 \right\}^{1/2}, \min_{s_k \ge 2} d^+(\caly_k) \right], 
    \end{equation*} 
    which completes the proof.
\end{proof}

\subsection*{Proof of Theorem~\ref{thm:fcodd}}
\begin{proof}
We first show that 
  \begin{align*}
  d^*(\bar \gamma) &= w_k/(s_k-1) = \max \{d^*(\gamma): \gamma_i \in \calG_k, m(\gamma)=s_k\} ,\\
  d^+(\bar \gamma) &= 2w_k/(s_k-1) = \max \{d^+(\gamma): \gamma_i \in \calG_k, m(\gamma)=s_k\},
\end{align*}
and there is no other $\gamma$ that fulfills both properties. 

  On one hand,
  suppose there is a $\gamma \subset \calG_k$ 
such that $m(\gamma)=s_k$ and $d^*(\gamma) > w_k/(s_k-1)$. 
Then $\gamma_{(1)} \ge 0$ and $\gamma_{(z+1)} \ge \gamma_{(z)} + d^*(\gamma) > \gamma_{(z)} + w_k/(s_k-1) $ for any $z$. 
From an induction, we can show that $\gamma_{(s_k)} > w_k$, which conflicts to the assumption that $\gamma 
\subset \calG_k = [0,w_k]$. 
Therefore, $d^*(\bar \gamma) = w_k/(s_k-1) = \max \{d^*(\gamma): \gamma \subset \calG_k, m(\gamma)=s_k \} $.

  Similarly, suppose there is a $\gamma \subset \calG_k$ such that $m(\gamma)=s_k$ and $d^+(\gamma) > 2w_k/(s_k-1)$. 
Then $\gamma_{(1)} \ge 0$ and $\gamma_{(2z+1)} \ge \gamma_{(2z-1)} +d^+(\gamma) > \gamma_{(2z-1)} +2w_k/(s_k-1)$ for any $z$. 
From an induction, we can show that $\gamma_{(s_k)} > w_k$, which conflicts to the assumption that $\gamma 
\subset \calG_k = [0,w_k]$. 
Therefore, $d^+(\bar \gamma) = 2w_k/(s_k-1) = \max \{d^+(\gamma): \gamma \subset \calG_k, m(\gamma)=s_k \}$.

Finally, suppose there is a $\gamma \subset \calG_k$ 
such that $m(\gamma)=s_k$ and $d^*(\gamma) \geq w_k/(s_k-1)$. 
Then $\gamma_{(1)} \ge 0$ and for any $z$, $\gamma_{(z+1)} \ge \gamma_{(z)} + d^*(\gamma) \geq \gamma_{(z)} + w_k/(s_k-1) $ and the equality holds if and only if $\gamma_{(z+1)} = \gamma_{(z)} + w_k/(s_k-1) $. 
From an induction, we can show that $\gamma_{(s_k)} \geq w_k$ and the only choice that makes $\gamma_{(s_k)} = w_k$ is $\gamma = \bar\gamma$. 
Consequently, $\bar\gamma$ is the only admissible choice. 
\end{proof}

\subsection*{Proof of Theorem~\ref{thm:fceven}}
\begin{proof}
We first show that for any $0 \le v_1 \le w_k/(s_k-1)$, 
\begin{align*}
d^*(\tilde \gamma_{v_1})&= v_1 = \max \{d^*(\gamma): \gamma_i \in \calG_k, m(\gamma)=s_k, d^+(\gamma)\geq v_1+v_2\},\\
d^+(\tilde \gamma_{v_1}) &= v_1 + v_2 = \max \{d^+(\gamma): \gamma_i \in \calG_k, m(\gamma)=s_k, d^*(\gamma) \ge v_1 \}, 
\end{align*}
and $\tilde \gamma_{v_1}$ is the only choice that fulfills both properties. 

Clearly, when $0 \le v_1 \le w_k/(s_k-1)$, 
    \begin{align} \label{eq:proofy}
      \tilde \gamma_{v_1,i} - \tilde \gamma_{v_1,i-1} = 
     \left\{  
       \begin{array}{ll}  
       v_1, &\quad i \text{~is even}, \\  
       v_2, &\quad i \text{~is odd}.   
       \end{array}  
   \right. 
   \end{align}
   Therefore, $ d^+(\tilde \gamma_{v_1})= v_1 + v_2= 2(w_k-v_1)/(s_k-2)$ and $d^*(\tilde \gamma_{v_1}) = \min \{ v_1,v_2\}$. Since $v_2-v_1 = 2\{w_k - (s_k-1)v_1\}/(s_k-2)$, $v_1 \le w_k/(s_k-1)$, and $s_k > 2$, we have $v_2 \ge v_1$. Therefore, $d^*(\tilde \gamma_{v_1}) = v_1$. 
 
 
Suppose there is a $ \gamma \subset \calG_k$ such that $m( \gamma) =s_k$, $d^*( \gamma) \ge v_1$, and $ d^+( \gamma) > v_2$. 
Then $ \gamma_{(1)} \ge 0$, $ \gamma_{(i+1)} \ge  \gamma_{(i)} + v_1$, and $ \gamma_{(i+1)} >  \gamma_{(i)} + v_1+v_2$ for any $i$. 
Since $s_k$ is even, from an induction we have $ \gamma_{(s_k)} \ge (s_k/2-1) (v_1+v_2) + v_1 > w_k$, which conflicts to the assumption that $ \gamma \subset \calG_k = [0,w_k]$. Therefore, $ d^+(\tilde \gamma_{v_1}) = \max \{d^+(\gamma): \gamma \subset \calG_k, m(\gamma)=s_k, d^*(\gamma) \ge v_1\}$. 
Similarly, we can show that $ d^*(\tilde \gamma_{v_1}) = \max \{d^*(\gamma): \gamma \subset \calG_k, m(\gamma)=s_k, d^+(\gamma) \ge v_1+v_2\}$ and $\tilde \gamma_{v_1}$ is the only choice that fullfills both requirements. 
Consequently, $\tilde \gamma_{v_1}$ with $0 \le v_1 \le w_k/(s_k-1)$ is an admissible choice. 

Since the $d^*$ for elements of $\{ \tilde \gamma_{v_1} : 0 \leq v_1 \leq w_k/(s_k-1) \}$ range from the lowest possible $d^*$, i.e., 0, to the highest possible $d^*$, i.e., $w_k/(s_k-1)$, it follows that $\{ \tilde \gamma_{v_1} : 0 \leq v_1 \leq w_k/(s_k-1) \}$ constitutes a minimal sufficient set.  
\end{proof}

\subsection*{Proof of Theorem~\ref{thm:fo}}
\begin{proof}
It suffices to show that for any $\gamma \subset \calG_k$ such that $m(\gamma)=s_k$, there is a $\tilde \gamma \in \calF(\calG_k, s_k)$ such that  $d^*(\tilde \gamma) \ge d^*(\gamma)$ and $d^+(\tilde \gamma) \ge d^+(\gamma)$.

Suppose $\gamma \in \calG_k$ and $m(\gamma)=s_k$. 
From Line~\ref{alg:fksk_ordinal:dstar} of Algorithm~\ref{alg:fksk_ordinal} and initially the $\tilde d^+ =0$, there exists a $\tilde \gamma \in \calF(\calG_k,s_k)$ such that $d^*(\tilde \gamma) \geq d^*(\gamma)$. 
Let $\tilde \gamma$ be the one with the lowest $d^*$ among element of $\calF(\calG_k,s_k)$ whose $d^*$ is no less than $d^*(\gamma)$.

%
Supppose $d^+(\gamma) > d^+(\tilde \gamma)$. 
From Line~\ref{alg:fksk_ordinal:dplus} of Algorithm~\ref{alg:fksk_ordinal}, there exists a $\bar \gamma \in \calF(\calG_k,s_k)$ such that $d^*(\gamma) \leq d^*(\bar \gamma) < d^*(\tilde \gamma)$, which contradicts to the assumption that $\tilde \gamma$ be the one with the lowest $d^*$ among element of $\calF(\calG_k,s_k)$ whose $d^*$ is no less than $d^*(\gamma)$.
Consequently, the assumption that $d^+(\gamma) > d^+(\tilde \gamma)$ cannot be fulfilled. 
Therefore, 
$d^+(\tilde \gamma) \geq d^+(\gamma)$. 
\end{proof}

\section*{S2: Examples} 
\begin{example}
  Suppose $\calG=\{0,0.12,0.24, \allowbreak 0.36, \allowbreak 0.48,0.6,0.72,0.84,0.96,1\}$ and $s \in \{2,\ldots,9\}$,  Table~\ref{tab:fksk} presents the sets $\calF(\calG, s)$ computed using Algorithm~\ref{alg:fksk_ordinal}.

\begin{table}[t!]
  \caption{The $\calF(\calG,s)$ for $\calG = \{0,0.12,0.24,0.36,0.48,0.6,0.72,0.84,0.96,1\}$ and $s \leq 9$.}
  \label{tab:fksk}
  \centering
  \begin{tabular}{|cccc|}
    \hline
    $s$ & $\calF(\calG,s)$ & $d^*$ & $d^+$ \\ \hline
    2 & $\{0,1\}$ & 1 & $+\infty$ \\ \hline
    3 & $\{0,0.48,1\}$ & 0.48 & 1 \\ \hline
    4 & $\{0,0.36,0.72,1\}$ & 0.28 & 0.64 \\
      & $\{0,0.24,0.72,1\}$ & 0.24 & 0.72 \\
      & $\{0,0.24,0.84,1\}$ & 0.16 & 0.76 \\
      & $\{0,0.12,0.84,1\}$ & 0.12 & 0.84 \\ \hline
    5 & $\{0,0.24,0.48,0.72,1\}$ & 0.12 & 0.48 \\ \hline
    6 & $\{0,0.12,0.36,0.48,0.72,1\}$ & 0.24 & 0.36 \\ \hline
    7 & $\{0,0.12,0.36,0.48,0.72,0.84,1\}$ & 0.12 & 0.28 \\ \hline
    8 & $\{0,0.12,0.36,0.48,0.60,0.72,0.84,1\}$ & 0.12 & 0.24 \\ \hline
    9 & $\{0,0.12,0.24,0.36,0.48,0.60,0.72,0.84,1\}$ & 0.12 & 0.24 \\ \hline
  \end{tabular}
\end{table}

\end{example}

\begin{example} \label{exp:showalg1}
  Suppose $p=3$, $n=62$, and $\calG = \{0,0.12,0.24,0.36,0.48,0.6, \\0.72,0.84, \allowbreak 1\} \times [0,1]^2$. 
  The $\calF(\calG_1,s_1)$ for various $s_1$ has been shown in Table~\ref{tab:fksk}. 
There are 6 distinct standard interleaved lattices for $p=3$.  

\begin{align*}\label{eqn:G:33}
  &\bm{G}_{3,1} =  \begin{pmatrix}
  1&0&0\\
  0&1&0\\
  0&0&1
  \end{pmatrix}, 
     \quad \bm{G}_{3,2} = \begin{pmatrix}
                   1&0&1\\
                   0&1&1\\
                   0&0&2
             \end{pmatrix}, 
     \quad \bm{G}_{3,3} = \begin{pmatrix}
                       1&1&1\\
                       0&2&0\\
                       0&0&2
                 \end{pmatrix},
\\
&\bm{G}_{3,4} = \begin{pmatrix}
    1&0&0\\
    0&1&1\\
    0&0&2
    \end{pmatrix}, 
\quad \bm{G}_{3,5} = \begin{pmatrix}
        0&1&0\\
        1&0&1\\
        0&0&2
     \end{pmatrix},     
     \quad \bm{G}_{3,6} =  \begin{pmatrix}
                     0&0&1\\
                     1&1&0\\
                     0&2&0
         \end{pmatrix}, \nonumber 
\end{align*}
are the corresponding generator matrices.

  In below we illustrate the steps of Algorithm~\ref{alg:pless5}.
  In Line~\ref{alg6:initial}, we initialize $\hat\rho=0$, $\cals_1 = \{2,\ldots,9\}$, and $\cals_2=\cals_3 = \{2,\ldots,+\infty\}$. 
  For the first three-dimensional standard interleaved lattice, $\calL(\bm{G}_{3,1})$, we first select $\bm{s}=(2,2,16)$, the smallest $\bm{s}$ in lexicographical order that satisfies $m\{\calL(\bm{G}_{3,1}),\bm{s},\bm{0}_p\} \ge n$. 
  Using Algorithm~\ref{alg:Findy}, the optimal
  $\caly$ is found to be $\caly=\{0,1\}^2\times \{0,1/15,2/15,\ldots,1\}$, resulting in a separation distance of ${\rho}=0.067$.
  Because $0.067 > \hat\rho = 0$, in Lines~\ref{alg6:checkrho}-\ref{alg6:updateS} we update $\hat\rho=0.067$, $t=1$, $\hat \calL^{(1)} = \calL(\bm{G}_{3,1})$, $\bm{\hat s}^{(1)} = (2,2,16)$, $\hat \caly^{(1)} = \{0,1\}^2\times \{0,1/15,2/15,\ldots,1\}$,  
  $\cals_1 = \{2,\ldots,9\}$, and $\cals_2=\cals_3 = \{2,\ldots,31\}$. 
  We then proceed to test the second choice for $\bm{s}$, which is $(2, 3, 11)$.
  This choice leads to $\caly = \{0,1\}\times \{0,0.5,1\} \times \{0,1/10,2/10,\ldots,1\}$ and $\rho = 0.1$.
  Since $0.1 > 0.067 = \hat\rho$, in Lines~\ref{alg6:checkrho}-\ref{alg6:updateS} we update $\hat\rho=0.1$, $t=1$, $\hat \calL^{(1)} = \calL(\bm{G}_{3,1})$, $\bm{\hat s}^{(1)} = (2,3,11)$, $\hat \caly^{(1)} = \{0,0.5,1\} \times \{0,1/10,2/10,\ldots,1\}$, $\cals_1 = \{2,\ldots,9\}$, and $\cals_2=\cals_3 = \{2,\ldots,21\}$. 
  As we explore more combinations of $\calL$ and $\bm{s}$, the value of $\hat{\rho}$ gradually increases. 
  Table~\ref{tab:lsys} presents all combinations of $\calL$, $\bm{s}$, and $\caly$ that either satisfy $\rho\{\calD(\calL,\bm{s},\caly,\bm{0}_p)\} > \hat{\rho} $ or $\rho\{\calD(\calL,\bm{s},\caly,\bm{0}_p)\} = \hat{\rho} $, along with the corresponding updated values of  $\hat\rho$ and $\cals$.

\begin{table}[htbp]
    \scriptsize
      \caption{The updates of $\hat\rho$, $\calL$, ${\bm{s}}$, $\caly$, and $\cals$ from Algorithm~\ref{alg:pless5} for Example~\ref{exp:showalg1}.}
      \centering
      \begin{tabular}{|p{7mm}p{8mm}p{8mm}p{70mm}p{18mm}|}
        \hline   
        $\hat\rho$&$\calL$ & ${\bm{s}}$ &$\caly$& $\cals$ \\ \hline
        \multirow{2 }{=}{0.067} & \multirow{2 }{=}{$\calL(\bm{G}_{3,1})$} & $\multirow{2 }{=}{(2,2,16)}$&\multirow{2 }{=}{$\{0,1\}^2 \times \{0,1/15,2/15,\ldots,1\}$} & $ \{2,\ldots,9\} $\\
        &&&&$ \times  \{2,\ldots,31\}^2$\\ 
        \hline
        \multirow{2 }{=}{0.100} & \multirow{2 }{=}{$\calL(\bm{G}_{3,1})$} & $\multirow{2 }{=}{(2,3,11)}$&\multirow{2 }{=}{$\{0,1\}\times \{0,0.5,1\} \times \{0,1/10,2/10,\ldots,1\}$} & $ \{2,\ldots,9\}$ \\ 
        &&&& $ \times  \{2,\ldots,21\}^2$\\ \hline 
        \multirow{2 }{=}{0.143} & \multirow{2 }{=}{$\calL(\bm{G}_{3,1})$} & $\multirow{2 }{=}{(2,4,8)}$&\multirow{2 }{=}{$\{0,1\}\times \{0,1/3,2/3,1\}\times\{0,1/7,2/7,\ldots,1\}$} & $ \{2,\ldots,9\}$\\ 
        &&&& $\times  \{2,\ldots,15\}^2$\\ \hline 
        \multirow{2 }{=}{0.167} & \multirow{2 }{=}{$\calL(\bm{G}_{3,1})$} & $\multirow{2 }{=}{(2,5,7)}$&\multirow{2 }{=}{$\{0,1\}\times \{0,1/4,2/4,1\}\times\{0,1/6,2/6,\ldots,1\}$} & $ \{2,\ldots,9\} $\\ 
        &&&&$ \times  \{2,\ldots,13\}^2$\\ \hline 
        \multirow{2 }{=}{0.200} & \multirow{1 }{=}{$\calL(\bm{G}_{3,1})$} & $\multirow{1 }{=}{(2,6,6)}$&\multirow{1 }{=}{$\{0,1\}\times \{0,1/5,2/5,\ldots,1\}^2$} & $ \{2,\ldots,9\} $\\   
         & \multirow{1 }{=}{$\calL(\bm{G}_{3,1})$} & $\multirow{1 }{=}{(3,4,6)}$&\multirow{1 }{=}{$ \{0,0.48,1\}\times  \{0,1/3,2/3,1\}\times  \{0,1/5,2/5,\ldots,1\}$} &$ \times  \{2,\ldots,11\}^2$\\  
          \hline   
         \multirow{2 }{=}{0.250} & \multirow{2 }{=}{$\calL(\bm{G}_{3,1})$} & $\multirow{2 }{=}{(3,5,5)}$&\multirow{2 }{=}{$\{0,0.48,1\}\times \{0,1/4,2/4,3/4,1\}^2$} & $ \{2,\ldots,7\}$\\ 
         &&&& $  \times  \{2,\ldots,9\}^2$\\ \hline  
         \multirow{6 }{=}{0.280} & \multirow{1 }{=}{$\calL(\bm{G}_{3,1})$} & $\multirow{1 }{=}{(4,4,4)}$&\multirow{1 }{=}{$\{0,0.36,0.72,1\}\times \{0,1/3,2/3,1\}^2$} &\\   
         & \multirow{1 }{=}{$\calL(\bm{G}_{3,3})$} & $\multirow{1 }{=}{(4,4,8)}$&\multirow{1 }{=}{$\{0,0.36,0.72,1\}\times  \{0,1/3,2/3,1\}\times  \{0,1/7,\ldots,1\}$} &\\ 
         & \multirow{1 }{=}{$\calL(\bm{G}_{3,3})$} & $\multirow{1 }{=}{(4,5,7)}$&\multirow{1 }{=}{$\{0,0.36,0.72,1\}\times  \{0,1/4,\ldots,1\}\times  \{0,1/6,\ldots,1\}$} & $ \{2,\ldots,7\}$\\    
         & \multirow{1 }{=}{$\calL(\bm{G}_{3,3})$} & $\multirow{1 }{=}{(4,6,6)}$&\multirow{1 }{=}{$\{0,0.36,0.72,1\}\times  \{0,1/5,2/5,\ldots,1\}^2$} & $ \times  \{2,\ldots,8\}^2$\\
         & \multirow{1 }{=}{$\calL(\bm{G}_{3,3})$} & $\multirow{1 }{=}{(4,7,5)}$&\multirow{1 }{=}{$\{0,0.36,0.72,1\}\times \{0,1/6,\ldots,1\} \times \{0,1/4,\ldots,1\}$} &\\  
         & \multirow{1 }{=}{$\calL(\bm{G}_{3,3})$} & $\multirow{1 }{=}{(4,8,4)}$&\multirow{1 }{=}{$\{0,0.36,0.72,1\}\times \{0,1/7,\ldots,1\} \times \{0,1/3,2/3,1\}$} &\\  \hline   
         \multirow{3 }{=}{0.299} & \multirow{1 }{=}{$\calL(\bm{G}_{3,5})$} & $\multirow{1 }{=}{(4,4,8)}$&\multirow{1 }{=}{$\{0,0.36,0.72,1\}\times \{0,1/3,2/3,1\} \times \{0,1/7,\ldots,1\}$} & $ \{2,\ldots,6\} $\\   
         & \multirow{1 }{=}{$\calL(\bm{G}_{3,6})$} & $\multirow{1 }{=}{(4,8,4)}$&\multirow{1 }{=}{$\{0,0.36,0.72,1\}\times\{0,1/7,\ldots,1\} \times \{0,1/3,2/3,1\}$} &$ \times  \{2,\ldots,8\}^2$\\ 
         & \multirow{1 }{=}{$\calL(\bm{G}_{3,2})$} & $\multirow{1 }{=}{(4,4,8)}$&\multirow{1 }{=}{$\{0,0.36,0.72,1\}\times \{0,1/3,2/3,1\}\times \{0,1/7,\ldots,1\}$} &\\  \hline   
         \multirow{2 }{=}{0.300} & \multirow{1 }{=}{$\calL(\bm{G}_{3,2})$} & $\multirow{1 }{=}{(4,5,7)}$&\multirow{1 }{=}{$\{0,0.36,0.72,1\}\times \{0,1/4,\ldots,1\}\times \{0,1/6,\ldots,1\}$} &$ \{2,\ldots,6\} $\\   
         & \multirow{1 }{=}{$\calL(\bm{G}_{3,2})$} & $\multirow{1 }{=}{(4,6,6)}$&\multirow{1 }{=}{$ \{0,0.36,0.72,1\}\times  \{0,1/5,2/5,\ldots,1\}^2$} &$  \times  \{2,\ldots,8\}^2$\\    \hline   
         \multirow{2 }{=}{0.347} & \multirow{2 }{=}{$\calL(\bm{G}_{3,2})$} & $\multirow{2 }{=}{(5,5,5)}$&\multirow{2 }{=}{$\{0,0.24,0.48,0.72,1\}\times \{0,1/4,2/4,3/4,1\}^2$} & \multirow{2 }{=}{$\{2,\ldots,6\}^3$}\\ 
         &&&&\\ \hline 
      \end{tabular}
      \label {tab:lsys}
  \end{table}

  At the end of Line~\ref{alg6:endfor} we obtain $t=1$, $\hat \calL^{(1)} = \calL(\bm{G}_{3,2})$, $\bm\hat{s}^{(1)}=(5,5,5)$, $\hat \caly^{(1)} =\{0,0.24,0.48,0.72,1\} \times \{0,0.25,0.5,0.75, 1\}^2 $, and $\hat{\rho}=0.3466$. 
  Thus, for the case with $p=3$, $n=62$, and $\calG = \{0,0.12,0.24,0.36,0.48,0.6,0.72,0.84, \allowbreak 1\} \times [0,1]^2$, the highest achievable separation distance from interleaved lattice-based maximin distance designs for mixed-type variables is $0.3466$, which is attained by only one $(\calL, \bm{s}, \caly)$. 
  The corresponding design, $\calD(\hat \calL^{(1)}, \bm\hat{s}^{(1)},\hat \caly^{(1)},\bm{0}_p)$, which consists of $m=63$ points, is depicted in Fig.~\ref{fig:example}(c). 
  In Line~\ref{alg6:findbest}, we determine that one of the optimal translations is $\bm{u}=(0,0,1)$, resulting in a design with $m=62$ points, as shown in Fig.~\ref{fig:example}(d). 
  As a result, the final solution obtained from Algorithm~\ref{alg:pless5} is $\hat{\calL} = \calL(\bm{G}_{3,2})$, $\hat{\bm{s}}=(5,5,5)$, $\hat\caly=\{0,0.24,0.48,0.72,1\} \times \{0,0.25,0.5,0.75, 1\}^2$,  $\bm{\hat{u}}=(0,0,1)$, $\hat{\rho}=0.347$, and $\hat{m}=62$. 
\end{example}

\begin{spacing}{1.7}
  \bibhang=1.7pc
  \bibsep=2pt
  \fontsize{9}{14pt plus.8pt minus .6pt}\selectfont
  \renewcommand\bibname{\large \bf References}
  \expandafter\ifx\csname
  natexlab\endcsname\relax\def\natexlab#1{#1}\fi
  \expandafter\ifx\csname url\endcsname\relax
    \def\url#1{\texttt{#1}}\fi
  \expandafter\ifx\csname urlprefix\endcsname\relax\def\urlprefix{URL}\fi

    \bibliographystyle{chicago}      
  \bibliography{MmM_ordinal_Bio}

  \vskip .2cm
  \noindent
  School of Mathematics, Statistics and Mechanics, Beijing University of Technology,  Pingleyuan No. 100, Chaoyang District, Beijing, 100124, China.
  \vskip 2pt
  \noindent
  E-mail: huilan@bjut.edu.cn
  \vskip 2pt
  
  \noindent
  State Key Laboratory of Mathematical Sciences (SKLMS), Academy of Mathematics and Systems Science, Chinese Academy of Sciences,  Haidian District, Beijing 100190, China.
  \vskip 2pt
  \noindent
  E-mail: hexu@amss.ac.cn
  
  \end{spacing}

\end{document}